	\documentclass{article}
	\usepackage{latexcad}
\begin{document}
	\baselineskip18truept									
\begin{center}
\LARGE\scshape Non-Abelian Stokes theorem in action
\end{center}

\begin{center}
{\Large\mdseries Bogus{\l}aw Broda}%
\footnote{e-mail: {\tt bobroda@krysia.uni.lodz.pl}}
\end{center}

\begin{center}
\large\textit{
Department of Theoretical Physics\\
University of \L\'od\'z\\
Pomorska 149/153\\
PL 90--236 \L\'od\'z\\
Poland}
\end{center}


\begin{abstract}
In this short review main issues related to the non-Abelian Stokes theorem
have been addressed. The two principal approaches to the non-Abelian Stokes
theorem, operator and two variants (coherent-state and holomorphic) of the
path-integral one, have been formulated in their simplest possible forms. A
recent generalization for a knotted loop as well as a suggestion concerning
higher-degree forms have been also included. Non-perturbative applications
of the non-Abelian Stokes theorem, to \hbox{(semi-)to}\-pological gauge theories,
have been presented.
\end{abstract}


\section{Introduction}
For tens years the (standard, i.e.\ Abelian) {\it Stokes} theorem is one of the central points of
(multivariable) analysis on manifolds. Lower-dimensional versions of this
theorem, known as the (proper) Stokes theorem, in
dimensions 1 and 2, and the, so-called {\it Gauss} theorem, in
dimensions 2 and 3, respectively, are well-known and extremally
useful in practice, e.g.\ in classical electrodynamics (Maxwell equations).  In fact, it is
difficult, if not impossible, to imagine lectures on classical electrodynamics
without intensive use of the Stokes theorem.  The standard
Stokes theorem is also being called the {\it Abelian}
Stokes theorem, as it applies to  (ordinary, i.e.\ Abelian)
differential forms. Classical electrodynamics is an Abelian
(i.e.\ $U(1)$) gauge field theory (gauge fields are Abelian forms), therefore
its integral formulas are governed by the Abelian Stokes
theorem. But much of interesting and physically important 
phenomena is described by non-Abelian gauge theories.
Hence it would be very desirable  to
have at our disposal a non-Abelian version of the Stokes
theorem. Since non-Abelian differential forms need
 a bit different treatment, one is forced to use a more
sophisticated formalism to deal with this new situation.

The aim of this chapter is to present a short review of the
non-Abelian Stokes theorem. At first, we will give an account of different formulations of the  non-Abelian Stokes theorem and next of various applications of thereof.

\subsection{Abelian Stokes theorem}
Before we engage in the  non-Abelian Stokes theorem it seems reasonable to recall its Abelian version. The (Abelian) Stokes theorem says  (see, e.g.\ [Spi65], for an excellent introduction to the subject) that 
we can convert an integral around a closed curve
$C$ bounding some surface $S$ into an integral
defined on this surface. Namely, in e.g.\ three dimensions
	\begin{equation}
	\oint_C \vec{A} \cdot d\vec{s}
	=\int_S {\rm curl}\vec{A} \cdot
	\vec{n}\,d\sigma, 
	\label{prop.Stok.}
	\end{equation}
\begin{figure}[htb]
\begin{center}
\begin{picture}(58,60)
\thinlines
\drawshadedcircle{20.0}{20.0}{72.0}{$S$}{0.1}
\drawvector{56.0}{18.0}{2.0}{0}{1}
\drawcenteredtext{-30.0}{40.0}{$\partial S=C$}
\end{picture}
\end{center}
\caption{Integration areas for the lowest-dimensional (non-trivial) version of the Abelian Stokes theorem.}
\label{fig:ast}
\end{figure}
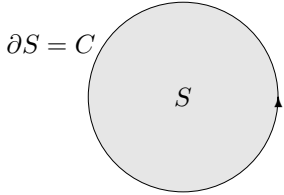
where the curve $C$ is the boundary 
of the surface $S$, i.e.\ $C=\partial S$ (see, Fig.\ref{fig:ast}),
$\vec{A}$ is a vector field, e.g.\ the vector 
potential of electromagnetic field,
and $\vec{n}$ is a unit outward normal at 
the area element $d\sigma$.

More generally, in any dimension,
	\begin{equation}
	\int_{\partial N} \omega=\int_N d\omega ,
	\label{ab.Stok.}
	\end{equation}
where now $N$ is a $d$-dimensional submanifold of the manifold $M$,
$\partial N$ is its $(d-1)$-dimensional 
boundary, $\omega $ is a $(d-1)$-form,
and $d\omega $ is its differential, a $d$-form.
We can also rewrite Eq.\ref{prop.Stok.} in the spirit of Eq.\ref{ab.Stok.}, i.e.
	\begin{equation}
	\oint_{\partial S=C} A_i\, dx^i=
	\frac {1}{2} \int_S
	\left( \partial_i A_j-\partial_j A_i \right) dx^i \wedge
	dx^j ,
	\label{ab.prop.Stok.} 
	\end{equation}
where $A_i$ ($i=1,2,3$) are components of the vector $\vec A$, 
and the Einstein summation convention after repeating indices is assumed.

In electrodynamics, we define the stress
tensor of electromagnetic field
$$
F_{ij}=\partial_i A_j -\partial_j A_i , 
$$
and the magnetic induction, its dual, as
$$
B_k=\frac {1}{2} \varepsilon_{ijk}F_{ij}, 
$$
where $\varepsilon_{ijk}$ is the totally antisymmetric
(pseudo-)tensor. 
RHS of Eq.\ref{ab.prop.Stok.} represents then the magnetic flux through $S$.
Thus, we can rewrite Eq.\ref{ab.prop.Stok.} in the form of \ref{prop.Stok.}
\begin{eqnarray}
	\oint_{\partial S=C} A_i\, dx^i
	&=&	\frac {1}{2} \int_S F_{ij}\, dx^i \wedge dx^j
	\label{ab.prop.Stok.2}\\
	&=&\int_S B_in^i\, d\sigma.
	\nonumber
\end{eqnarray}

In turn, in geometry $\vec A$ plays the role of  connection 
(it defines the parallel transport around $C$), and $F$ is
the curvature of this connection.
A ``global version'' of the Abelian Stokes theorem,
	\begin{equation}
	\exp \left( i \oint_{\partial S=C}
	A_i(x)\, dx^i \right) =
	\exp \left( \frac {i}{2} 
	\int_S F_{ij}(x)\, dx^i \wedge dx^j \right),
	\label{glob.ab.Stok.}
	\end{equation}
which is rather a trivial generalization of Eq.\ref{ab.prop.Stok.2},  is a very good starting point 
for our further discussion 
concerning the non-Abelian Stokes theorem.
The object on LHS of \ref{glob.ab.Stok.} is called the holonomy, and more generally, for open curves $C$, global connection.

\subsection{Historical remarks}
The birth of the ideas related to the (non-)Abelian Stokes theorem dates back to the ninetenth century and is connected with the emergence of the Abelian Stokes theorem. The Abelian Stokes theorem can be treated as a prototype of the  non-Abelian Stokes theorem or a version of thereof when we confine ourselves to an Abelian group.

A  work closer to the proper non-Abelian Stokes theorem, by Schlesinger [Sch27], where generally non-commuting matrix-values functions have been considered appeared in 1927. In fact no  work on the genuine  non-Abelian Stokes theorem could appear before the birth of very idea of non-Abelian gauge fields in the beginning of fifthies. And really, first papers on the  non-Abelian Stokes theorem appeared in the very end of seventhies. At first, the  non-Abelian Stokes theorem emerged in the operator version [Hal79] (the first appearance of the non-Abelian Stokes theorem, see Eq.3.8 therein), [Are80], [Bra80], [FGK81], and later on, in the very end of eighties, in the path-integral one [DP89], [Bro92].

\subsection{Contents}
The propert part of the paper consists of two sections. The first section is devoted to the non-Abelian Stokes theorem itself. In the beginning, we introduce necessary notions and conventions. The operator version of the non-Abelian Stokes theorem is formulated in the first subsection. The second subsection concerns the path-integral versions of the non-Abelian Stokes theorem: coherent-state approach and holomorphic approach. The last subsection describes generalizations of the non-Abelian Stokes theorem. First of all, to topologically more general situations, and also to higher-degree forms. The second section is devoted to applications of the non-Abelian Stokes theorem in mathematical and theoretical physics. In the first subsection an approach to the computation of Wilson loops in two-dimensional Yang-Mills theory is presented. The second subsection deals with the analogous problem for three-dimensional (topological) Chern-Simons gauge theory. Other possibilities, including higher-dimensional gauge theories and QCD are mentioned in the last subsection.

\section{Non-Abelian Stokes theorem}
What is the non-Abelian Stokes theorem? To answer this question we should recall first of all the form of the well-known Abelian Stokes theorem. Namely (see, Eq.\ref{ab.Stok.})
	\begin{equation}
	\int_{\partial M} \omega = \int_M d\omega,
	\label{ab.Stok.2}
	\end{equation}
where the integral of the form $\omega$ along the boundary $\partial N$ of the submanifold $N$ is equated to the integral of the differential $d\omega$ of this form over the submanifold $N$. The differential form $\omega$ is usually an ordinary (i.e.\ Abelian) differential form, but it could also be something more general, e.g.\ connection one-form $A$. Thus, the  non-Abelian Stokes theorem should be a version of \ref{ab.Stok.2} for non-Abelian (say, Lie-algebra valued) forms. Since it could be risky to directly integrate Lie-algebra valued differential forms, the generalization of \ref{ab.Stok.2} may be a non-trivial task. We should not be too ambitious perhaps from the very beginning, and not try to formulate the  non-Abelian Stokes theorem in full generality at once, as it could be difficult or even impossible simply. The lowest non-trivial dimensionality of the objects entering \ref{ab.Stok.2} is as follows: $\dim N=2$ ($\dim\partial N=1$) and $\deg\omega=1$ ($\deg d\omega=2$). A short reflection leads us to the first candidate for the LHS of the  non-Abelian Stokes theorem, the Wilson loop
	\begin{equation}
	{\rm P}\exp\left(i\oint_C A\right),
	\label{Pexp}
	\end{equation}
called the holonomy, in mathematical context, where $\rm P$ denotes the, so-called, path ordering, $A$ is a non-Abelian connection one-form, and $C$ is a closed loop, a boundary of the surface $S$ ($\partial S=C$). Correspondingly, the RHS of the  non-Abelian Stokes theorem should contain a kind of integration over $S$. Therefore, the actual Abelian prototype of the  non-Abelian Stokes theorem is of the form \ref{glob.ab.Stok.} rather than of \ref{ab.prop.Stok.2}. More often, the trace of Eq.\ref{Pexp} is called the Wilson loop,
\begin{equation}
	W_R(C)
	=
	{\rm Tr}_R{\rm P}\exp\left(i\oint_C A\right),
	\label{TrPexp}
\end{equation}
or even the ``normalized'' trace of it,
	\begin{equation}
	\frac{1}{\dim R}{\rm Tr}_R{\rm P}
	\exp\left(i\oint_C 	A\right),
	\label{norm.TrPexp}
	\end{equation}
where the character $R$ means a(n irreducible) representation of the Lie group $G$ corresponding to the given Lie algebra $\bf g$. Of course, one can easily pass from \ref{Pexp} to \ref{TrPexp} and finally to \ref{norm.TrPexp}. In fact, the operator \ref{Pexp} is a particular case of a more general parallel-transport operator
	\begin{equation}
	U_L={\rm P}\exp\left(i\int_L A\right),
	\label{par.trans.}
	\end{equation}
where $L$ is a smooth path, which for the $L$ a closed loop ($L=C$) yields \ref{Pexp}. Eq.\ref{par.trans.} could be considered as an ancestor of Eq.\ref{Pexp}. As the LHS of the  non-Abelian Stokes theorem we can assume any of the formulas given above for the closed loop $C$ (i.e. Eqs.\ref{Pexp}, \ref{TrPexp}, \ref{norm.TrPexp}) yielding possibly various versions of the  non-Abelian Stokes theorem. For some reason or another, sometimes it is more convenient to use the Wilson loop in the operator version \ref{Pexp} rather than in the version with the trace \ref{TrPexp}. Sometimes it does not make any bigger difference. The RHS should be some expression defined on the surface $S$, and essentially constituting the  non-Abelian Stokes theorem.

From the point of view of a physicist the expression \ref{par.trans.} is typical in quantum mechanics and corresponds to the evolution operator. By the way, Eqs.\ref{TrPexp}, \ref{norm.TrPexp} are very typical in gauge theory, e.g.\ in QCD. Thus, guided by our intuition we can reformulate our chief problem as a quantum-mechanical one. In other words, the approaches to the LHS of the  non-Abelian Stokes theorem are analogous to the approaches to the evolution operator in quantum mechanics. There are the two main approaches to quantum mechanics, and especially to the construction of the evolution operator: opearator approach and path-integral approach. The both can be applied to the  non-Abelian Stokes theorem successfully, and the both provide two different formulations of the  non-Abelian Stokes theorem.

%
\paragraph{Conventions}
Sometimes, especially in a physical context, a coupling constant, denoted e.g.\ $e$, appears in front of the integral in Eqs.\ref{Pexp}--\ref{par.trans.}. For simplicity, we will omit the coupling constant in our formulas.

The non-Abelian
curvature or the strength field on the manifold $M$ is defined by
$$
F_{ij}(A)
=\partial_i A_j-\partial_jA_i-i[A_i,A_j].
$$
Here, the connection or the gauge potential $A$ assuming values
in a(n 
irreducible) representation $R$ of the compact, semisimple Lie algebra $\bf g$ of the Lie group $G$
is of the form
	$$
	A_i(x)=A_i^a(x) T^a,
	\qquad
	i=1,\ldots,\dim{M},
	$$
where the Hermitian generators, $T^{a\dag}=T^a$, $T^a=T^a_{kl}$,
$k,l=1,\ldots,\dim R$, fulfil the commutation relations
\begin{equation}	
	[T^a,T^b]=if^{abc}T^c,
	\qquad
	a,b,c=1,\ldots,\dim{G}.
	\label{Liealg.com.}
\end{equation}
The line integral \ref{par.trans.} can be rewritten in more detailed (being frequently used in our further analysis) forms,
\[
U(x'',x')={\rm P}\exp\left[i\int_{x'}^{x''} \!\!\!A_i(x)\,dx^i\right],
\]
or
\[   
U_{kl}= {\rm P}\exp\left[i\int_{x'}^{x''}
\!\!\!A_i^a(x) T^a\,dx^i\right],
\]
without parametrization, and
\[
U(t'',t')={\rm P}\exp\left\{i\int_{x'}^{x''} A_i^a[x(t)] T^a \frac{dx^i(t)}{dt}\,dt\right\},
\]
with an explicit parametrization, or some variations of thereof. Here, the oriented smooth path $L$ starting at the point $x'$ and endng at the point $x''$ is parametrized by the function $x^i(t)$, where $t'\leq t\leq t''$, and $x^{\prime i}=x^i(t')$, $x^{\prime\prime i}=x^i(t'')$.

\subsection{Operator formalism}
Unfortunately, it is not possible to automatically 
generalize the Abelian Stokes theorem (e.g.\ Eq.\ref{ab.prop.Stok.2}) to the
non-Abelian one. In the non-Abelian case one faces a
qualitatively 
different situation because
the integrand on the LHS assumes  values in  a Lie
algebra  $\bf g$ rather 
than in  the field of real or complex numbers.
The picture simplifies significantly 
if one switches from the ``local'' language to a global one (see, Eq.\ref{glob.ab.Stok.}). 
Therefore we should consider the holonomy \ref{Pexp} around 
a closed curve $C$,
$$
{\rm P}\exp 
\left( i\oint_C A_i dx^i \right) . 
$$
The holonomy represents a parallel-transport
operator 
around $C$ assuming values in a non-Abelian Lie group
$G$. 
(Interestingly, in the Abelian case, the holonomy has a physical , it is an object playing the role of the
phase which can be observed in the Aharonov-Bohm
experiment, whereas $A_i$ itself has not such an interpretation.)

%
\paragraph{Non-Abelian Stokes theorem}
The
non-Abelian generalization of Eq.\ref{glob.ab.Stok.} should read
$$
{\rm P} \exp \left( i\oint_{\partial S=C}
A_i(x) dx^i \right)
= {\cal P} \exp \left( \frac{i}{2}
\int_S {\cal F}_{ij} (x) dx^i\wedge dx^j \right),
$$
where the LHS has been already roughly defined. As far as the RHS is concerned, the symbol $\cal P$ denotes
some ``surface ordering'', whereas ${\cal F}_{ij} (x)$ is a 
``path-dependent curvature'' given by the formula
$$
{\cal F}_{ij} (x)
\stackrel{\rm def}{=}
 U^{-1}(x,O) F_{ij}(x) U(x,O),
$$
where $U(x,O)$ is a parallel-transport operator along
the path $L$ in the surface $S$ joining
\placedrawing{oval.fig}{Parallel-transport operator along the path $L$ in the surface $S$.}{fig:oval}
the base point $O$ of $\partial S$ with the point
$x$, i.e.
$$
U(x,O)={\rm P} \exp \left( i\int_L
A_i(y)dy^i \right) .
$$
See, Fig.\ref{fig:oval}, and further sections for more details.

\subsubsection{Calculus of paths}%
	\label{SECTION:calculusofPaths}
As a kind of a short introduction for properly manipulating parallel-transport operators along oriented curves we recall a number of standard facts. It is obvious that we can perform some operations on the parallel transport operators. We can superpose them, we can introduce an identity element, and finally we can find an inverse element for each element. 

Roughly, the structure is similar to the structure of the group with the following standard postulates satisfied:
(1) associativity, $(U_1U_2)U_3=U_1(U_2U_3)$; (2) existence of an identity element, $IU=UI=U$; (3) existence of an inverse element $U^{-1}$, $U^{-1}U=UU^{-1}=I$.
\begin{figure}[htb]
\begin{center}
\hbox to\textwidth{\vtop{\hsize=.5\textwidth%
\advance\hsize by -.5\columnsep
\parindent=0pt
\centering
\begin{picture}(28,60)
\thinlines
\drawvector{4.0}{4.0}{20.0}{0}{1}
\drawvector{4.0}{24.0}{0.0}{0}{1}
\drawvector{4.0}{24.0}{30.0}{0}{1}
\drawvector{24.0}{4.0}{50.0}{0}{1}
\drawcenteredtext{14.0}{54.0}{$x_1$}
\drawcenteredtext{14.0}{6.0}{$x_2$}
\drawcenteredtext{10.0}{22.0}{$x$}
\drawcenteredtext{14.0}{30.0}{$=$}
\end{picture}
\caption{Allowable composition of elements.}
\label{fig:comp.}
\vskip1sp}\hskip\columnsep\vtop{\hsize=.5\textwidth%
\advance\hsize by -.5\columnsep
\parindent=0pt
\centering
\begin{picture}(30,56)
\thinlines
\drawvector{6.0}{6.0}{40.0}{0}{1}
\drawvector{26.0}{46.0}{40.0}{0}{-1}
\drawcenteredtext{18.0}{26.0}{$=$}
\drawcenteredtext{12.0}{46.0}{}
\drawleftbrace{3.0}{26.0}{10.0}
\drawcenteredtext{18.0}{6.0}{$x_2$}
\drawcenteredtext{14.0}{48.0}{}
\drawcenteredtext{18.0}{46.0}{$x_1$}
\drawrightbrace{9.0}{26.0}{10.0}
\drawcenteredtext{12.0}{50.0}{-1}
\end{picture}
\caption{Inverse element.}
\label{fig:inv.}
}\hfill}
\end{center}
\end{figure}
But let us note that not all elements can be superposed. Although parallel-transport operators are elements of a Lie group $G$ but their geometrical interpretation has been lost in the notation above. We can superpose two elements only when the end point of the first element is the initial point of the second one. Thus, $U_1U_2$ could be meaningfull in the form (Fig.\ref{fig:comp.})
\[
U(x_1,x)U(x,x_2)=U(x_1,x_2).
\]
Obviously,
\[
I=U(x,x),
\]
and (Fig.\ref{fig:inv.})
\[
U^{-1}(x_1,x_2)=U(x_2,x_1).
\]
The above formulas become a particularly convincing in a graphical form.
Perhaps, one of the most useful facts is expressed by the following Fig.\ref{fig:comp.}:
\begin{figure}[ht]
	\begin{center}
	\begin{picture}(30,68)
\thinlines
\drawvector{4.0}{4.0}{50.0}{0}{1}
\drawvector{24.0}{4.0}{20.0}{0}{1}
\drawvector{24.0}{24.0}{20.0}{1}{0}
\drawvector{44.0}{24.0}{2.0}{0}{1}
\drawvector{44.0}{26.0}{20.0}{-1}{0}
\drawvector{24.0}{26.0}{28.0}{0}{1}
\drawcenteredtext{14.0}{28.0}{$=$}
\drawcenteredtext{15.0}{6.0}{$x_2$}
\drawcenteredtext{15.0}{52.0}{$x_1$}
\drawcenteredtext{50.0}{20.0}{$x$}
\end{picture}	
	\end{center}
	\caption{Deformation of a path.}
	\label{fig:def.}
\end{figure}
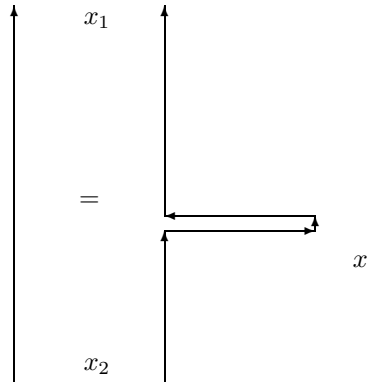

It appears that this structure fits into the structure of the, so-called, grouppoid.

	\subsubsection{Ordering}
There are a lot of different ordering operators in our formulas which have been collected in this section.

%
\paragraph{Dyson series}
We know from quantum theory that the path-ordered exponent of an operator $\hat{A}$ can be expressed by the power series called the Dyson series,
\begin{eqnarray}
		\lefteqn{{\rm P}\,\exp\left[i\int_{t'}^{t''}\!\hat{A}(t)\;dt\right]}   \nonumber \\
	&=& {\rm P}\left[1+\sum_{n=1}^{\infty}\frac{i^n}{n!}
	\int_{t'}^{t''}\!dt_1\int_{t'}^{t''}\!dt_2\ldots
	\int_{t'}^{t''}\!dt_n\,\hat{A}(t_1)\hat{A}(t_2)\ldots
	\hat{A}(t_n)\right]   \nonumber \\
	&=&1+\sum_{n=1}^{\infty}\frac{i^n}{n!}
	\int_{t'}^{t''}\!dt_1\int_{t'}^{t''}\!dt_2\ldots
	\int_{t'}^{t''}\!dt_n\,
	{\rm P}\left[\hat{A}(t_1)\hat{A}(t_2)\ldots
	\hat{A}(t_n)\right]   \nonumber \\
	&=&1+\sum_{n=1}^{\infty}i^n
	\int_{t'}^{t''}\!dt_1\int_{t'}^{t_1}\!dt_2\ldots
	\int_{t'}^{t_{n-1}}\!dt_n\,
	\hat{A}(t_1)\hat{A}(t_2)\ldots\hat{A}(t_n),
	\nonumber\end{eqnarray}
where
\[
{\rm P}\left[\hat{A}_1(t_1)\ldots\hat{A}_n(t_n)\right]
\stackrel{\rm def}{=}
\hat{A}_{\sigma(1)}\left(t_{\sigma(1)}\right)\ldots
\hat{A}_{\sigma(n)}\left(t_{\sigma(n)}\right)),
\qquad
t_{\sigma(1)}\geq\ldots\geq t_{\sigma(n)}.
\]
For example, for two operators
\[
	{\rm P}\left[\hat{A}_1(t_1)\hat{A}_2(t_2)\right]
	\equiv \theta\left(t_1-t_2\right)\hat{A}_1(t_1)\hat{A}_2(t_2)
	+\theta\left(t_2-t_1\right)\hat{A}_2(t_2)\hat{A}_1(t_1),
	 	\]
where $\theta$ is the step function.

%
\paragraph{Conventions}
Since the operators/matrices appearing in our considerations are, in general, non-commutative, we assume the following conventions:
\begin{equation}
	\prod_{n=1}^N X_n
	\stackrel{\rm def}{=}
	X_N X_{N-1} \ldots X_2 X_1,
	\nonumber
\end{equation}
\begin{equation}
	{\rm P}_t \left(X_1 X_2 \ldots X_{N-1} X_N\right)
	\stackrel{\rm def}{=}
	X_N X_{N-1} \ldots X_2 X_1,
	\nonumber
\end{equation}
whereas for two parameters
\begin{eqnarray}
	\lefteqn{{\rm P}_{s,t}\prod_{m,n=1}^N X_{m,n}}   \nonumber \\
&\stackrel{\rm def}{=}&
	\prod_{n=1}^N \prod_{m=1}^N X_{m,n}\nonumber\\	
	&\equiv&
	X_{N,N} X_{N-1,N} \ldots X_{1,N} X_{N,N-1} \ldots X_{1,2} X_{N,1} \ldots X_{2,1} X_{1,1}.    \nonumber
\end{eqnarray}

\subsubsection{Theorem}
The  non-Abelian Stokes theorem in its original operator form roughly claims that the holonomy round a closed curve $C=\partial S$ equals a surface-ordered exponent of the {\it twisted} curvature, namely
\begin{equation}
		{\rm P}\exp\left(i\oint_C A \right)
		={\cal P}\exp\left(i\int_S {\cal F}\right),
	\label{op.NAST}
\end{equation}
where $\cal F$ is the twisted curvature 
\[
	{\cal F}\equiv U^{-1} F U.
\]
A more precise form of the  non-Abelian Stokes theorem as well as an exact meaning of the notions appearing in the theorem will be given in the course of the proof.

%
\paragraph{Proof}
Following [Are80] and [Men83] we will present a short, direct proof of the  non-Abelian Stokes theorem.

In our parametrization,
\placedrawing{bigsq.fig}{The parametrized loop $C$ as a boundary of the ``big'' square $S$ with the coordinates
$\{(0,0),(1,0),(1,1),(0,1)\}$.}{fig:bigsq}
the first step consists of the decomposition of the initial loop (see, Fig.\ref{fig:bigsq}) into small lassos according to the rules given in the
section \ref{SECTION:calculusofPaths}
\[
	{\rm P}\exp\left(i\oint_C A \right)
	=\lim_{N\rightarrow\infty}({\rm P}_{s,t})\prod_{m,n=1}^N U_{m,n}^{-1} W_{m,n} U_{m,n},
\]
where the objects involved are defined as follows:
parallel-transport operators from the reference point
\placedrawing{umn.fig}{Approaching the ``small'' plaquette inside the ``big'' square.}{fig:umn}
($0,0$) to the point with coordinates $(\frac{m}{N},\frac{n}{N})$ consists of two segments
(see, Fig.\ref{fig:umn})
\[
	U_{m,n}
	\stackrel{\rm def}{=}
	{\rm P}\exp\left(i\int_{(0,\frac{n}{N})}^%
	{(\frac{m}{N},\frac{n}{N})}A\right)
	{\rm P}\exp\left(i\int_{(0,0)}^%
	{(0,\frac{n}{N})}A\right),
	\]
parallel-transport operator round a small plaquette $S_{m,n}$
\placedrawing{smallsq.fig}{The small plaquette.}{fig:smallsq}
	\[
	W_{m,n}
	\stackrel{\rm def}{=}
	 	{\rm Tr\,P}\exp\left(i\oint_{S_{m,n}} A \right),
\]
where $S_{m,n}$ is a boundary of a (small) square with coordinates
	\[
	S_{m,n}:\quad\left\{
\left(\frac{m}{N},\frac{n}{N}\right), 
\left(\frac{m-1}{N},\frac{n}{N}\right), 
\left(\frac{m-1}{N},\frac{n-1}{N}\right), 
\left(\frac{m}{N},\frac{n-1}{N}\right) 
\right\}.
\]
Now
\begin{eqnarray}
	U_{m,n}^{-1} W_{m,n} U_{m,n}
	&=& U_{m,n}^{-1} \left[1+\frac{i}{N^2}F_{m,n}
	+O\left(\frac{1}{N^3}\right)\right] U_{m,n}	   
	\nonumber \\
	&=& 1+\frac{i}{N^2} U_{m,n}^{-1} F_{m,n} U_{m,n}	 
	+O\left(\frac{1}{N^3}\right) 	   
	\nonumber \\
	&=& 1+\frac{i}{N^2}{\cal F}_{m,n}
	+O\left(\frac{1}{N^3}\right)   
	\nonumber \\
	&=& {\cal W}_{m,n}+O\left(\frac{1}{N^3}\right),
\nonumber
\end{eqnarray}
where $W_{m,n}$ has been calculated in the Appendix to this section (see, Eq.\ref{Wmn}),
	\[
	{\cal F}_{m,n}
	\stackrel{\rm def}{=}
	 U_{m,n}^{-1} F_{m,n} U_{m,n},
	\]
and
	\[
	{\cal W}_{m,n}
	\stackrel{\rm def}{=}
	\exp\left(\frac{i}{N^2}{\cal F}_{m,n}\right).
\]
Then 
\begin{eqnarray}
	{\rm P}\exp\left(i\oint_C A \right)
	&=&
	\lim_{N\rightarrow\infty}({\rm P}_{s,t})\prod_{m,n=1}^N
	\left[{\cal W}_{m,n}+O\left(\frac{1}{N^3}\right)\right]
	\nonumber \\
	&=&
	\lim_{N\rightarrow\infty}({\rm P}_{s,t})\prod_{m,n=1}^N
	\exp\left[\frac{i}{N^2}{\cal F}_{m,n}
	+O\left(\frac{1}{N^3}\right)\right]\nonumber \\
	&=&
	\lim_{N\rightarrow\infty}({\rm P}_t)\prod_{m,n=1}^N
	\exp\left[\frac{i}{N^2}{\cal F}_{m,n}
	+O\left(\frac{1}{N^3}\right)\right].
	\nonumber
\end{eqnarray}
The last equality follows from the fact that operations corresponding to the change of the order of the operators yield the commutator
	\[
	\left[\frac{1}{N^2}{\cal F}_{m',n'},
	\frac{1}{N^2}{\cal F}_{m'',n''}\right]
	=O\left(\frac{1}{N^4}\right),
\]
and there is maximum $N-1$ transpositions possible in the framework of $s$-ordering, so
	\[
	(N-1)\,O\left(\frac{1}{N^4}\right)
	=O\left(\frac{1}{N^3}\right).
\]
Thus, we arive at the final form of the  non-Abelian Stokes theorem,
\[
	{\rm P}\exp\left(i\oint_C A \right)
	=\lim_{N\rightarrow\infty}({\rm P}_t)\prod_{m,n=1}^N
	\exp\left[\frac{i}{N^2}{\cal F}_{m,n}\right]
	={\cal P}\exp\left(i\int_S {\cal F}\right).
	\]

%
\paragraph{Appendix}
In this paragraph we will perform a fairly standard calculus and derive the contribution coming from a small loop $W_{m,n}$. Namely,
	\begin{eqnarray}
		W_{m,n}
			&=&
			{\rm P}\exp\left(i\int%
_{\left(\frac{m}{N},\frac{n-1}{N}\right)}
^{\left(\frac{m}{N},\frac{n}{N}\right)}
		A\right)
		{\rm P}\exp\left(i\int%
_{\left(\frac{m-1}{N},\frac{n-1}{N}\right)}
^{\left(\frac{m}{N},\frac{n-1}{N}\right)}
		A\right)
		\nonumber
		\\
		&&\quad{\rm P}\exp\left(i\int%
_{\left(\frac{m}{N},\frac{n-1}{N}\right)}
^{\left(\frac{m-1}{N},\frac{n-1}{N}\right)}
		A\right)
		{\rm P}\exp\left(i\int%
_{\left(\frac{m}{N},\frac{n}{N}\right)}
^{\left(\frac{m}{N},\frac{n-1}{N}\right)}
		A\right)\nonumber\\			 
	&=&
	\left[1+\frac{i}{N}A_{m,n}^y
	+\frac{1}{2}\left(\frac{i}{N}A_{m,n}^y\right)^2
	+O\left(\frac{1}{N^3}\right)\right]
	\nonumber\\
	&&\quad	
	\cdot\left[1+\frac{i}{N}A_{m,n-1}^x
	+\frac{1}{2}\left(\frac{i}{N}A_{m,n-1}^x\right)^2
	+O\left(\frac{1}{N^3}\right)\right]
	\nonumber\\
	&&\quad	
	\cdot\left[1-\frac{i}{N}A_{m-1,n}^y
	+\frac{1}{2}\left(\frac{i}{N}A_{m-1,n}^y\right)^2
	+O\left(\frac{1}{N^3}\right)\right]
	\nonumber\\
	&&\quad		
	\cdot\left[1-\frac{i}{N}A_{m,n}^x
	+\frac{1}{2}\left(\frac{i}{N}A_{m,n}^x\right)^2
	+O\left(\frac{1}{N^3}\right)\right]	
	\nonumber\\
	&=& 
	1+\frac{i}{N}\left(A_{m,n}^y - A_{m-1,n}^y\right)
	-\frac{i}{N}\left(A_{m,n}^x - A_{m,n-1}^x\right)
	\nonumber\\
	&&\quad
	+\left(\frac{i}{N}\right)^2 \left(A_{m,n}^y\right)^2
	+\left(\frac{i}{N}\right)^2 \left(A_{m,n}^x\right)^2
	+\left(\frac{i}{N}\right)^2 A_{m,n}^y A_{m,n}^x
	\nonumber\\
	&&\quad
	-\left(\frac{i}{N}\right)^2 \left(A_{m,n}^y\right)^2
	-\left(\frac{i}{N}\right)^2 A_{m,n}^y A_{m,n}^x
	-\left(\frac{i}{N}\right)^2 A_{m,n}^x A_{m,n}^y\nonumber\\
	&&\quad	
	-\left(\frac{i}{N}\right)^2 \left(A_{m,n}^x\right)^2
	+\left(-\frac{i}{N}\right)^2 A_{m,n}^y A_{m,n}^x
	+O\left(\frac{1}{N^3}\right)\nonumber\\
	&=&
	1+\frac{i}{N^2}\left(\partial_xA^y-\partial_yA^x\right)
	-\frac{i^2}{N^2}\left(A^xA^y-A^yA^x\right)
	+O\left(\frac{1}{N^3}\right)\nonumber\\
	&=&
	1+\frac{i}{N^2}F_{m,n}
	+O\left(\frac{1}{N^3}\right),\label{Wmn}
			\end{eqnarray}
where
\[
	F=\partial_xA_y-\partial_yA_x-i\left[A_x,A_y\right]
	\]
(ocassionally, we put $x$ and $y$ in the upper position), and
\[
	A=A_{m,n}+O\left(\frac{1}{N}\right).
	\]
Then
\[
	W_{m,n}=\exp\left[\frac{i}{N^2}F_{m,n}
	+O\left(\frac{1}{N^3}\right)\right].
	\]

%
\paragraph{Other approaches}
There is a lot of other approaches to the (operator) non-Abelian Stokes theorem, more or less interrelated. The analytic approach advocated in [Bra80] and [HM98]. The approach using the, so-called, product integration [KMR99] and [KMR00]. And last, but not least, a (very interesting) coordinate-gauge approach [SS98], and [Hal79] as well as its lattice formulation [Bat82].

\subsection{Path-integral formalism}
There are two main approaches to the  non-Abelian Stokes theorem in the framework of the path-integral formalism: coherent-state approach and holomorphic approach. In the literature, the both approaches occur in few, a bit different, incarnations. Also, the both have found applications in different areas of mathematical and/or theoretical physics, and therefore the both are usefull. The first one is formulated more in the spirit of group theory, whereas the second one follows from traditional path-integral formulation of quantum mechanics or rather quantum field theory. Similarly to the situation in quantum theory, the path-integral formalism is easier in some applications and more intuitive then the operator formalism but traditionally it is mathematically less rigorous.
In the same manner as quantum mechanics, initially
formulated in the operator language and next reformulated
into the path-integral one, we can translate the operator
form of the non-Abelian Stokes theorem into the
path-integral one.

In order to
formulate the non-Abelian Stokes theorem in the
path-integral language we will make the following three
steps:
\begin{enumerate}

\item
We will determine a coherent-state/holomorphic path-integral 
representation for the parallel-transport operator
deriving an appropriate transition amplitude (a
path-integral counterpart of the LHS in Eq.\ref{op.NAST});

\item
For a closed curve $C$ we will calculate the trace of the path-integral form of the parallel-transport operator,
quantum theory in an external gauge field $A$;

\item
We will apply the Abelian
Stokes theorem to the exponent of the integrand of the path integral yielding the RHS of the  non-Abelian Stokes theorem (a counterpart of  RHS in Eq.~2.8).
\end{enumerate}

%
\paragraph{Preliminary formulas}
Since the both path-integral derivations of the transition amplitude have a common starting point that is independent of the particular approach, we present it here:
\begin{equation}
	{\rm P}\exp\left[i\int_{t'}^{t''}A(t)dt\right]=
	\lim_{N\rightarrow\infty}\prod_{n=1}^{N}
	\left(1+i\epsilon A_n\right),
	\label{prel.form.}
\end{equation}
where
\begin{equation}
	A_n=A(t_n),\qquad
	\epsilon=\frac{t''-t'}{N},\qquad
	t_n=n\epsilon+t',\qquad
	t_N=t'',\quad t_0=t'.
	\nonumber	
\end{equation}
From this moment the both approaches differ.

\subsubsection{Coherent-state approach}
%
%
\paragraph{Group-theoretic coherent states}
According to [ZFG90] (see, also [Per86]) the group-theoretic coherent states emerge in the following construction:

\begin{enumerate}
\item
For $\bf g$, a semisimple Lie algebra of a Lie group $G$, we introduce the standard Cartan basis $\{H_i,E_\alpha,E_{-\alpha}\}$:
\begin{eqnarray}
	\left[H_i,H_j\right]&=&0,
	\nonumber\\
	\left[H_i,E_\alpha\right]&=& \alpha_iE_\alpha,
	\nonumber\\	
	\left[E_\alpha,E_{-\alpha}\right]&=&\alpha^iH_i,
	\nonumber\\
	\left[E_\alpha,E_\beta\right]
	&=&N_{\alpha;\beta}E_{\alpha+\beta}.
	\label{Car.bas.}
	\end{eqnarray}

\item
We chose a unitary irreducible representation $R$ of the group $G$, as well as a normalized state the, so-called, reference state $\left|R\right>$. The choice of the reference state is in principle arbitrary but not unessential. Usually it is an ``extremal state'' (the highest weight state), the state anihilated by $E_\alpha$, i.e.\ $E_\alpha\left|R\right>=0$.

\item
A subgroup of $G$ that consists of all the group elements $h$ that will leave the reference state $\left|R\right>$ invariant up to a phase factor is the maximum-stability subgroup $H$. Formally,
\[
	h\left|R\right\rangle
	=
	\left|R\right\rangle e^{i\phi(h)},
	\qquad h\in H.
\]
The phase factor is unimportant here because we shall generally take the expectation value of any operator in the coherent state.

\item
For every element $g\in G$, there is a unique decomposition of $g$ into a product of two group elements, one in $H$ and the other in the quotient $G/H$:
\[
	g=\xi h,\quad g\in G,\quad h\in H,\quad \xi\in G/H.
\]
In other words, we can obtain a unique coset space for a given $\left|R\right\rangle$.

\item
One can see that the action of an arbitrary group element $g\in G$ on $\left|R\right\rangle$ is given by
\[
	g\left|R\right\rangle=
	\xi h\left|R\right\rangle=
	\xi\left|R\right\rangle e^{i\phi(h)}.
\]
The combination
\[
\left|\xi,R\right\rangle
\stackrel{\rm def}{=}
   \xi\left|R\right\rangle
   \]
   is the general group definition of the coherent states.
For simplicity, we will denote the coherent states as $\left|g,R\right\rangle$.
\end{enumerate}
The coherent states $\left|g,R\right\rangle$ are generally non-orthogonal but normalized to unity,
\[
\left<g,R\right.\left|g,R\right>=1.
\]
Furthermore, for an appropriately normalized measure $d\mu(g)$, we have a very important for our furher analysis identity the, so-called, resolution of unity
\begin{equation}
	\int\left|g,R\right>d\mu(g)\left<g,R\right|=I.
	\label{res.uni.}
\end{equation}

%
\paragraph{Path integral}
Our first aim is to calculate the ``transition amplitude'' between the two coherent states $\left|g',R\right\rangle$ and $\left|g'',R\right\rangle$
\begin{eqnarray}
	&&\left\langle g'',R\left|
{\rm P}\exp\left[i\int_{t'}^{t''}\!\!\!A(t)\,dt\right] \right|g',R\right\rangle  
\nonumber \\
	&&=\lim_{N\rightarrow\infty}	\int\ldots\int
\left\langle g_N,R\left|\left(1+i\epsilon A_N\right) \right|g_{N-1},R\right\rangle d\mu(g_{N-1})    \nonumber \\
	&&\qquad\left\langle g_{N-1},R\left|\left(1+i\epsilon A_{N-1}\right) \right|g_{N-2},R\right\rangle d\mu(g_{N-2})    \nonumber \\
	&&\qquad\ldots d\mu(g_1)\left\langle g_1,R\left|\left(1+i\epsilon A_1\right) \right|g_0,R\right\rangle,
\label{coh.pathint.}
\end{eqnarray}
where we have used \ref{prel.form.} and \ref{res.uni.}.
To continue, one should evaluate a single amplitude (i.e.\ the amplitude for an infinitesimal ``time'' $\epsilon$), i.e.
\begin{eqnarray}
	&&\left\langle g_n,R\left|\left(1+i\epsilon A_n\right) 	\right|g_{n-1},R\right\rangle    \nonumber \\	
	&&=\left\langle g_n,R | g_{n-1},R\right\rangle
+i\left\langle g_n,R\left| A_n \right|g_{n-1},R\right\rangle\epsilon.
\nonumber
\end{eqnarray}
Now
\begin{eqnarray}
\left\langle g_n,R | g_{n-1},R\right\rangle
	&=&\left\langle g(t_n),R|g(t_{n-1}),R\right\rangle    \nonumber \\
	&=&\left\langle R \left|g^\dagger(t_n) g(t_{n-1})\right|R\right\rangle
	=\left\langle R \left|g^\dagger(t_n) g(t_n-\epsilon)\right|R\right\rangle   \nonumber \\
	&=&\left\langle R \left|g^\dagger(t_n) \left[g(t_n)-\dot{g}(t_n)\epsilon+O\left(\epsilon^2\right)\right]\right|R\right\rangle   \nonumber \\
	&=&\left\langle R \left|g^\dagger(t_n) g(t_n)\right|R\right\rangle
-\left\langle R \left|g^\dagger(t_n) \dot{g}(t_n)\right|R\right\rangle\epsilon
+O\left(\epsilon^2\right)   \nonumber \\
	&=&1-	\left\langle R \left|g^\dagger(t_n) \dot{g}(t_n)\right|R\right\rangle\epsilon
+O\left(\epsilon^2\right),
\nonumber
\end{eqnarray}
whereas
\begin{eqnarray}
	i\left\langle g_n,R\left| A_n \right|g_{n-1},R\right\rangle\epsilon
	&=&i\left\langle g(t_n),R\left| A(t_n) \right|g(t_{n-1}),R\right\rangle\epsilon   \nonumber \\
		&=&i\left\langle R\left|g^\dagger(t_n) A(t_n) g(t_n-\epsilon)\right|R\right\rangle\epsilon   \nonumber \\
&=&i\left\langle R\left|g^\dagger(t_n) A(t_n) g(t_n)\right|R\right\rangle\epsilon
+O\left(\epsilon^2\right).		
\nonumber
\end{eqnarray}
Then
\begin{eqnarray}
		\lefteqn{\left\langle g_n,R\left|\left(1+i\epsilon A_n\right) 	\right|g_{n-1},R\right\rangle}   \nonumber \\
	&=&1-	\left\langle R \left|g^\dagger(t_n) \dot{g}(t_n)\right|R\right\rangle\epsilon
	+i\left\langle R\left|g^\dagger(t_n) A(t_n) g(t_n)\right|R\right\rangle\epsilon
+O\left(\epsilon^2\right)   \nonumber \\
	&=&\exp{\left[\left\langle R \left|-g^\dagger(t_n) \dot{g}(t_n)
	+ig^\dagger(t_n) A(t_n) g(t_n)\right|R\right\rangle\epsilon
	+O\left(\epsilon^2\right)\right]}.
\nonumber
\end{eqnarray}
Returning to \ref{coh.pathint.}
\[
\left\langle g'',R\left|
{\rm P}\exp\left[i\int_{t'}^{t''}\!\!\!A(t)\,dt\right] \right|g',R\right\rangle  
	=\int \left[D\mu(g)\right]
	\exp{\left(i\int_{t'}^{t''}\!\!\!L\, dt\right)},
\]
where the ``lagrangian'' appearing in the path integral is defined as
\begin{equation}
		L=\left<R \left|\left[ig^\dagger(t) \dot{g}(t)
		+g^\dagger(t) A(t) g(t)\right]\right|R\right>  
		\stackrel{\rm def}{=}
		\left<R \left|A^g(t)\right|R\right>,
		\label{coh.L}
\end{equation}
and
\[
\left[D\mu(g)\right]=\prod_{t'<t<t''} d\mu[g(t)].
\]

According to [HU00], wee can transform $L$ in Eq.\ref{coh.L} to the form originally proposed by [DP89]. Namely, for any Lie-algebra element $K$ we have in the Cartan basis \ref{Car.bas.}
\[
	\left<R \left|K\right|R\right>
	=
	\left<R \left|\sum_i \lambda_i H_i
	+\ldots\right|R\right>   
	=
	\sum_i \lambda_i\left<R \left| H_i 
	\right|R\right>,
	\]
where dots mean $E_{\mp\alpha}$ generators vanishing between $\left|R\right>$'s.
Since
\[
H_i\left|R\right>=m_i\left|R\right>,
\]
then
\begin{eqnarray}
\sum_i \lambda_i\left<R \left| H_i 
	\right|R\right>
	&=&
	\sum_i \lambda_i\left<R \left| R_i
	\right|R\right>
	=
	\sum_i \lambda_i m_i
	\nonumber\\
	&=&
	\frac{1}{\kappa}\sum_i m_i {\rm Tr}(H_iK)   
	=
	\frac{1}{\kappa} {\rm Tr}(\underline{m}
	\cdot \underline{H}\,K),
	\nonumber	
\end{eqnarray}
where the normalization
\[
{\rm Tr}(H_i H_j)=\kappa\delta_{ij}
\]
has been assumed.
Thus,
\begin{eqnarray}
	L&=&\left<R \left| A^g(t)\right|R\right> \nonumber\\
	&=&\frac{1}{\kappa}\sum_i{\rm Tr}
	\left[m_iH_iA^g(t)\right]\nonumber\\
	&=&\frac{1}{\kappa}{\rm Tr}
	\left\{\underline{m}\cdot\underline{H}
	\left[ig^\dagger(t) \dot{g}(t)
		+g^\dagger(t) A(t) g(t)\right]\right\}.
\end{eqnarray}

	\paragraph{Non-Abelian Stokes theorem}
Finally, the LHS of the  non-Abelian Stokes theorem reads
\[
\int D\mu(g)\exp{\left(i\oint_{t'}^{t''}\!\!\!L\,
dt\right)},
\]
where
\[
D\mu(g)=\prod_{t'\leq t<t''} d\mu[g(t)],
\]
and $g(t')=g(t'')$. Or, in the language of differential forms
\[
\int D\mu(g)\exp{\left(i\oint_C\!\!\!L\right)},
\]
where now
\[
D\mu(g)=\prod_{x\in C} d\mu[g(x)],
\]
and
\[
L=\frac{1}{\kappa}\sum_i{\rm Tr}
	\left[\underline{m}\cdot\underline{H}A_i^g(x)\right]dx^i
\stackrel{\rm def}{=}
B_i\,dx^i
\]
with
\[
A_i^g(x)=ig^\dagger(x) \partial_i g(x)
		+g^\dagger(x) A_i(x) g(x),
\]
\[
A^g(t)=A_i^g[x(t)]\frac{dx^i(t)}{dt}.
\]
Here $B_i$ is an Abelian differential form, so obviously
\[
	\int D\mu(g)\exp{\left(i\oint_{C=\partial S}
	\!\!\! B\right)}
 =\int D\mu(g)\exp{\left(i\int_S dB\right)}.
\]

\subsubsection{Holomorphic approach}
%
%
\paragraph{Quantum-mechanics background}
For our further convenience let us formulate
an auxiliary ``Schr\" odinger problem''
governing the parallel-transport operator \ref{par.trans.} for the Abelian gauge potential $A$,
\begin{equation}	
	i\frac {d z}{dt }=-\dot{x}^i A_i  z,
	\label{ab.Schr.eq.}
\end{equation}
which expresses the fact that  the ``wave function'' $ z$
should be covariantly constant along the line $L$, i.e.
$$
D_t   z
\stackrel{\rm def}{=}
 \left( \frac {d}{dt } -i\dot{x}^i A_i
\right)  z =0,
$$
where $D_t $ is the absolute covariant derivative.

First of all, let us derive the path-integral
expression for the parallel-transport operator $U$
along $L$. To this end, 
we should consider 
the non-Abelian formula (differential equation) 
analogous to Eq.\ref{ab.Schr.eq.},
\begin{equation}
	i\frac{d z_{k}}{dt }=-\dot{x}^i(t  ) A_i^a
	[x(t )] T^a_{kl} z_l,
	\label{nab.Schr.eq.}
\end{equation}
or
\[
(D_t   z)_k
\stackrel{\rm def}{=}
 \left\{\delta_{kl}\frac{d}{dt} -i\dot{x}^i
 A_i^a[x(t)] T^a_{kl}
\right\}  z_l =0,
\]
where $ z$ is an auxiliary ``wave function'' in an irreducible
representation $R$ of the gauge Lie group $G$, which is
to be parallelly transported along $L$ 
parametrized by $x^i(t  )$, 
$t  '\le t  \le t  ''$. Formally, Eq.\ref{nab.Schr.eq.} can
be instantaneously integrated out yielding
$$
 z_k (x'')=U_{kl}(x'',x') z_l (x') ,
$$
where $x''=x(t  '')$, $x'=x(t  ')$, and
$$
U(x'',x')\equiv U(t  '', t  ')=
{\rm P} \exp \left( i\int_{t  '}^{t  ''}
\!\!\dot{x}^i (t  ) A_i[x(t  )]\, dt  \right),
$$
as expected.

Let us  now consider  the following auxiliary
classical-mechanics problem with the classical
Lagrangian
\begin{equation}
	L\left(\bar{ z},  z \right)=i \bar{ z} D_t   z.
	\label{clas.L}
\end{equation}
The equation of motion for $ z$ following from Eq.\ref{clas.L} reproduces Eq.\ref{nab.Schr.eq.}, and yields the classical Hamiltonian

\begin{equation}
	H=i\dot{x}^i(t)A_i^a[x(t)] 
	T_{kl}^a \pi_k  z_l
	=-\dot{x}^i (t)A_i^a[x(t)]T_{kl}^a
	\bar{ z}_k z_l .
	\label{clas.H}
\end{equation}
The corresponding auxiliary quantum-mechanics
problem is given,  
according to Eq.\ref{clas.H}, by the Schr\" odinger equation
\begin{equation}
	i\frac {d}{dt } | \Phi \rangle =\hat{H}(t )|\Phi
	\rangle,
	\label{Schr.eq.} 
\end{equation}
with
$$
\hat{H}(t)
=-\hat{A}(t)
=-\dot{x}^i (t)A_i^a [x(t)]\hat{T}^a
=-\dot{x}^i(t) A_i^a[x(t)]T_{kl}^a\hat{a}_k^+\hat{a}_l 
\equiv H_{kl}(t  )
\hat{a}_k^+ \hat{a}_l,
$$
where the creation and anihilation operators satisfy the
standard commutation ($-$) or anticommutation ($+$)
relations
$$
[\hat{a}_k ,\hat{a}_l^+]_{\mp}= \delta_{kl} ,\quad
[\hat{a}_k^+ ,\hat{a}_l^+ ]_{\mp}
=[\hat{a}_k ,\hat{a}_l ]_{\mp} =0 .
$$
It can be easily checked by direct computation 
that we have really obtained a realization of the Lie algebra
$\bf g$ in a Hilbert (Fock) space,
$$
[ \hat{T}^a,\hat{T}^b ]_-
=i f^{abc}\hat{T}^c ,
$$
in accordance with \ref{Liealg.com.},
where $\hat{T}^a=T_{kl}^a \hat{a}_k^+\hat{a}_l$. For an 
irreducible representation $R$
the second-order Casimir operator $C_2$ is proportional
to the identity operator $I$, which in turn, is equal
to the number operator $\hat{N}$ in our Fock representation,
i.e.\ if $T^a\rightarrow \hat{T}^a$, then $I\rightarrow \hat{N} 
=\delta_{kl} \hat{a}_k^+ \hat{a}_l$. Thus, we obtain an important for our further considerations constant
of motion $\hat{N}$,
\begin{equation}
	[\hat{N},\hat{H} ]_- =0 .
	\label{[N,H]}
\end{equation}
It is interesting to note that this approach works equally well for commutation relations as well as for anticommutation relations.

%
\paragraph{Path integral}
Let us now derive the holomorphic path-integral
representation for the kernel of the parallel-transport operator,
\begin{eqnarray}
	\lefteqn{
	\left<\bar{z}''\left|{\rm P}\exp
	\left[i\int_{t'}^{t''}\!\!\!\hat{A}(t)\,dt\right]
	\right|z'\right>}
	\nonumber\\
	&=&
	\lim_{N\rightarrow\infty}\int\ldots\int\
	\left<\bar{z}_N\left|\left(1
	+i\epsilon \hat{A}_N\right)\right|z_{N-1}\right>
	e^{-\bar{z}_{N-1}z_{N-1}} 
	\frac{d\bar{z}_{N-1}\,dz_{N-1}}{2\pi i}
	\nonumber\\
	&&\quad\left<\bar{z}_{N-1}\left|\left(1
	+i\epsilon \hat{A}_{N-1}\right)\right|z_{N-2}\right>
	e^{-\bar{z}_{N-2}z_{N-2}} 
	\frac{d\bar{z}_{N-2}\,dz_{N-2}}{2\pi i}
	\nonumber\\
	&&\quad\ldots e^{-\bar{z}_1 z_1} 
	\frac{d\bar{z}_1\,dz_1}{2\pi i}
	\left<\bar{z}_1\left|\left(1
	+i\epsilon \hat{A}_1\right)\right|z_0\right>.
	\label{hol.amp.}
	\end{eqnarray}
Now we should calculate the single expectation value 
\[
	\left<\bar{z}_n\left|\left(1
	+i\epsilon \hat{A}_n\right)\right|z_{n-1}\right>
	=\left<\bar{z}_n\left|z_{n-1}\right.\right>
	+i\left<\bar{z}_n\left|\hat{A}_n
	\right|z_{n-1}\right>\epsilon.
	\]
Here
\[
	\left<\bar{z}_n\left|z_{n-1}\right.\right>
	=e^{\bar{z}_n z_{n-1}},
	\]
whereas
\begin{eqnarray}
	i\left<\bar{z}_n\left|\hat{A}_n
	\right|z_{n-1}\right>\epsilon
	&=&
	i\left<\bar{z}_n\left|\bar{z}_n A_n z_{n-1}
	\right|z_{n-1}\right>\epsilon   \nonumber \\
	&=&
	i\epsilon\bar{z}_n A_n z_{n-1}
	\left<\bar{z}_n\left|\right.z_{n-1}\right>.
	\nonumber
\end{eqnarray}
Thus
\begin{eqnarray}
	\left<\bar{z}_n\left|\left(1
	+i\epsilon \hat{A}_n\right)\right|z_{n-1}\right>
	&=&e^{\bar{z}_n z_{n-1}}
	+i\epsilon\bar{z}_n A_n z_{n-1}
	e^{\bar{z}_n z_{n-1}}  \nonumber \\
	&=&
	e^{\bar{z}_n z_{n-1}}\left[1
	+i\epsilon\bar{z}_n A_n z_n
	+O\left(\epsilon^2\right)\right]   \nonumber \\
	&=&
	e^{\bar{z}_n z_{n-1}}
	e^{+i\epsilon\bar{z}_n A_n z_n
	+O\left(\epsilon^2\right)}.
	\nonumber
\end{eqnarray}
Combining this expression with the exponent in \ref{hol.amp.} we obtain
\begin{eqnarray}
	e^{-\bar{z}_n z_n} e^{\bar{z}_n z_{n-1}
	+i\epsilon\bar{z}_n A_n z_n
	+O\left(\epsilon^2\right)}
 	&=&
	e^{-\bar{z}_n \left(z_n-z_{n-1}\right)
	+i\epsilon\bar{z}_n A_n z_n
	+O\left(\epsilon^2\right)}   \nonumber \\
	&=&
	\exp{\left[\left(-\bar{z}_n \dot{z}_n
	+\epsilon\bar{z}_n A_n z_n\right)\epsilon
	+O\left(\epsilon^2\right)\right]}.\nonumber
\end{eqnarray}
Finally, 
\begin{eqnarray}
	U(\bar{z}'',z';t'',t')
	&=&
	\left< \bar{z}'' \left| {\rm P} 
	\exp \left[-i\int_{t'}^{t''}\!\!\! 
	\hat{H}(t) dt  \right] \right| z' \right>
	\nonumber
	\\
	&\equiv&
	\left<\bar{z}''\left|{\rm P}\exp
	\left[i\int_{t'}^{t''}\!\!\!\hat{A}(t)\,dt\right]
	\right|z'\right>
	\nonumber
	\\
	&=&
	\int\left[D^2z\right]\exp\left\{\bar{z}(t'')z(t'')
	+i\int_{t'}^{t''}\!\left[i\bar{z}(t)\dot{z}(t)
	\right.\right.
	\nonumber
	\\
	&&\qquad
	\left.\left.
	+\bar{z}(t)A(t)z(t)\right]dt\right\},
	\nonumber
	\\
	&\equiv&
	\int\left[D^2z\right]\exp{\left[\bar{z}(t'')z(t'')
	+i\int_{t'}^{t''}\!\!\!L\,dt\right]},
	\nonumber
\end{eqnarray}
where
\[
	\left[D^2z\right]
	=\prod_{t'<t<t''} \frac{d\bar{z}(t)\,dz(t)}{2\pi i},
	\]
and $L$ is of ``classical'' form \ref{clas.L}.

Let us confine our attention to the one-particle
subspace of the Fock space. As the number operator $\hat{N}$
is conserved by  virtue of Eq.\ref{[N,H]}, if we start from the
one-particle subspace of the Fock space, we shall remain in
this subspace during all the evolution. The transition
amplitude $U_{kl}(t '',t ')$ between the one-particle
states $|1_k\rangle =\hat{a}_k^+ | 0\rangle$ and
$| 1_l\rangle = \hat{a}_l^+| 0\rangle$ is given by the
following scalar product in the holomorphic representation 
\begin{eqnarray}
	U_{kl}
	&=&
	\left<1_k\left|{\rm P}\exp	\left[i\int_{t'}^{t''}\!\!\!
	\hat{A}(t)\,dt\right]\right|1_l\right>
	\nonumber \\
	&=&
	\int\left<1_k\left|z''\right.\right>
	\left<\bar{z}''\left|{\rm P}
	\exp{\left[i\int_{t'}^{t''}\hat{A}(t)\,dt\right]}
	\right|z'\right>
	\left<\bar{z}'\left|1_l\right.\right>
	\nonumber \\
	&&\qquad
	e^{-\bar{z}''z''-\bar{z}'z'}
	\frac{d\bar{z}''\,dz''\,d\bar{z}'\,dz'}%
	{\left(2\pi i\right)^2} \nonumber  \\
	&=&\int D^2z\,z_k(t'')\bar{z}_l(t')
	\exp{\left[-\bar{z}(t')z(t')
	+i\int_{t'}^{t''}\!\!\!L\,dt\right]},
	\label{hol.par.tr.}
\end{eqnarray}
where now
\[
	D^2z
	=\prod_{t'\leq t\leq t''}
	\frac{d\bar{z}(t)\,dz(t)}{2\pi i}.
	\]
Depending 
on the  statistics, there are the two ($\mp $)
possibilities (fermionic and bosonic)
$$
z_k\bar{z}_l \mp \bar{z}_l z_k
=\bar{z}_k \bar{z}_l \mp \bar{z}_l \bar{z}_k
=z_k z_l\mp z_l z_k =0 ,
$$
equivalent as far as one-particle subspace of the Fock
space is concerned, which takes place 
in our further considerations.

One can easily check that Eq.\ref{hol.par.tr.} represents the object
we are looking for. Namely, from the Schr\"odinger
equation (Eq.\ref{Schr.eq.}) it follows that for the general
one-particle state $\alpha_k\hat{a}_k^+ | 0\rangle$
(summation after repeating indices) we have
\begin{equation}
	i\frac{d}{dt} (\alpha_m \hat{a}_m^+| 0\rangle )
	=H_{kl}(t)\hat{a}_k^+ \hat{a}_l \alpha_m \hat{a}_m^+
	| 0\rangle
	=H_{kl}(t  )\alpha_l (\hat{a}_k^+ | 0\rangle ).
	\label{Schr.1p} 
\end{equation}
Using the property of linear independence of 
Fock-space vectors in Eq.\ref{Schr.1p}, and comparing Eq.\ref{Schr.1p} to Eq.\ref{nab.Schr.eq.}, we can see that Eq.\ref{hol.par.tr.} really
represents the matrix elements of the
parallel-transport operator.
For
closed paths, $x(t')=x(t'')=x$, Eq.\ref{hol.par.tr.} gives the holonomy
operator $U_{kl}(x)$ and $U_{kk}$ is the Wilson
loop. Interestingly enough, the Wilson
loop, which is supposed to describe a quark-antiquark
interaction, is represented by a ``true'' quark and
antiquark field, $z$ and $\bar{z}$, respectively.  So, the
mathematical trick can be interpreted ``physically''.

Obviously, the ``full'' trace of the kernel in Eq.~3.9 is
obtained by imposing appropriate boundary conditions, and integrating with
respect to all the variables without the boundary term.
Analogously, one can also derive the parallel-transport
operator (a generalization of the one just considered) for
symmetric $n$-tensors (bosonic $n$-particle states) and for
$n$-forms (fermionic $n$-particle states).

%
\paragraph{The non-Abelian Stokes theorem}
Let us now define a (bosonic or fermionic) Euclidean
two-dimensional ``topological'' quantum field theory 
of multicomponent fields $\bar z, z$ transforming in an irreducible
representation $R$ of the Lie algebra $\bf g$
on the compact
surface $S$, $\mbox{dim} S=2, \partial S=C$, $S\subset M$,
$\dim{M}=d$, in an external non-Abelian gauge field $A$, by
the classical action
\begin{equation}
	S_{\rm{cl}}=\int_S \left( iD_i \bar{ z}D_j z +\frac{1}{2}
	\bar{ z} F_{ij} z \right) dx^i\wedge dx^j,
	\qquad i,j=1,\ldots ,d,
	\label{top.act.}
\end{equation}
or in a parametrization $x^i(\sigma^1,\sigma^2)$,
by the action	
\begin{eqnarray}
	S_{\rm{cl}}
	&=&
	\int_{S}{\cal L}_{\rm{cl}}(\bar{z},z)d^2\sigma
	\label{clas.eL}
	\\
	&=&
	\int_{S} \varepsilon^{AB}
	\left( i D_A\bar{ z} D_B z +\frac{1}{2} \bar{ z}
	F_{AB}z \right) d^2\sigma ,\qquad A,B=1,2,
	\nonumber
\end{eqnarray}
where
$$
D_A=\partial_A x^iD_i,
\qquad
D_i\stackrel{\rm def}{=}\partial_i-iA_i,
\qquad
F_{AB}=\partial_A x^i
\partial_B x^j F_{ij}.
$$

At present, we are prepared to formulate a holomorphic
path-integral version of the non-Abelian Stokes theorem,
\begin{eqnarray}
	\lefteqn{\int\! D^2z\,z_k''\bar{z}_k' 
	\exp{\left(-\bar{z}'z'
	+i\oint_C i\bar{z}\,Dz\right)}}
	\label{np.hol.NAST}
	\\
	&=&	 
	\int D^2z\,z_k''\bar{z}_k'
	\exp{\left(-\bar{z}'z'
	+iS_{\rm cl}\right)},
	\nonumber
	\end{eqnarray}
or in the polar parametrization
$x^i(\sigma^1,\sigma^2)$
$t  '\le \sigma^1 \equiv t  \le t  ''$,
$0\le \sigma^2\equiv s\le 1$,
\begin{eqnarray}
	\lefteqn{\int D^2z\,z_k(t'')\bar{z}_k(t') 
	\exp{\left\{-\bar{ z}(t') z(t')
	+i\int_{t'}^{t''}\!\!\!L[\bar{z}(t),z(t)]dt 
	\right\}}}
	\label{pp.hol.NAST}
	\\
	&=&	 
	\int D^2z\,z_k(t'',1)\bar{z}_k(t',1)
	\exp{\left[-\bar{z}(t',1)z(t',1)
	+i\int_0^1\!\!\!\int_{t'}^{t''}\!\!\!
	{\cal L}_{\rm{cl}}\,dt\,ds\right]},
	\nonumber
	\end{eqnarray}
where $L(z,\bar{z})$ and ${\cal L}_{\rm{cl}}(\bar{z},z)$
are defined by Eq.\ref{clas.L} and Eq.\ref{clas.eL}, respectively. The
measure on both sides of Eqs.\ref{np.hol.NAST} and \ref{pp.hol.NAST}  is the same, i.e. it is
concentrated on the boundary $\partial S$, and the imposed boundary conditions are free.

It should be noted that the surface
integral on the RHS of Eqs.\ref{np.hol.NAST} and \ref{pp.hol.NAST} depends on the curvature $F$
as well as on the connection $A$ entering the covariant
derivatives, which is reminiscent of the path dependence of
the curvature $\cal F$ in the operator approach.

A quite different formulation of the holomorphic approach to the non-Abelian Stokes theorem has been proposed in [Lun97].

\paragraph{Appendix}
For completness of our derivations, we will remind the reader a few standard facts being used above.
First of all, we assume the following (non-quite standard, but convinient) definition
\begin{eqnarray}
	\left.\left|z\right.\right>
	&\stackrel{\rm def}{=}&
	e^{za^++\bar{z}a+\frac{1}{2}|z|^2}
	\left.\left|0\right.\right>
	=e^{za^+} e^{\bar{z}a} \left.\left|0\right.\right>   \nonumber \\
	&=&e^{za^+} \left.\left|0\right.\right>
	=\sum_{k=0}^\infty \frac{\left(z^na^+\right)^n}{n!}
	\left.\left|0\right.\right>
	=\sum_{k=0}^\infty \frac{z^n}{\sqrt{n!}}
	\left.\left|n\right.\right>,
	\nonumber
\end{eqnarray}
where $\left.\left|0\right.\right>$ is the Fock vacuum, i.e. $\hat{a}\left.\left|0\right.\right>=0$, and the Baker-Campbell-Hausdorff formula has been applied in the first line. We could also treat $\left.\left|z\right.\right>$ as a coherent state for the Heisenberg group.
We can easily calculate
\[
	\left<\bar{z}\left|\right.z\right>
	=\sum_{m,n=0}^\infty \frac{\bar{z}^mz^n}{\sqrt{m!\,n!}}
	\left<m\left|\right.n\right>
	=\sum_{n=0}^\infty 
	\frac{\left(\bar{z}z\right)^n}{n!}
	=e^{\bar{z}z}.
	\]
The identity operator is of the form
\[
	I=\int\left.\left|z\right.\right>
	\left.\left<\bar{z}\right.\right|
	e^{-\bar{z}z} \frac{d\bar{z}\,dz}{2\pi i}.
	\]
Really,
\begin{eqnarray}
	I\left.\left|z'\right.\right>
	&=&\int\left.\left|z\right.\right>
	\left<\left.\bar{z}\right|z'\right>
	e^{-\bar{z}z} \frac{d\bar{z}\,dz}{2\pi i}
	=
	\int\left.\left|z\right.\right>
	e^{\bar{z}z'-\bar{z}z} \frac{d\bar{z}\,dz}{2\pi i}   \nonumber \\
	&=&
	\int\left.\left|z\right.\right>
	\delta(z'-z)\,dz=\left.\left|z'\right.\right>.
	\nonumber	
\end{eqnarray}
All these formulas for a single pair of creation and annihilation operators obviously apply to a more general situation of $\dim{R}$ pairs.	
The matrix elements are
\begin{eqnarray}
	\left<\bar{z}\left|\hat{A}\right|z\right>
	&=&
	\Big<0\Big|e^{\sum_k\bar{z}_k\hat{a}_k}
	\sum_a A^a \sum_{k,l} T_{kl}^a
	\hat{a}_k^+ \hat{a}_l e^{\sum_k z_k\hat{a}_k^+}
	\Big|0\Big>   \nonumber \\
	&=&
	\Big<0\Big|e^{\sum_k\bar{z}_k\hat{a}_k}
	\sum_a A^a \sum_{k,l} T_{kl}^a
	\bar{z}_k z_l e^{\sum_k z_k\hat{a}_k^+}
	\Big|0\Big>   \nonumber \\
	&=&
	\Big<\bar{z}\Big|
	\sum_a A^a \sum_{k,l} T_{kl}^a
	\bar{z}_k z_l \Big|z\Big>   \nonumber \\
	&=&\left<\bar{z}\left|\bar{z}Az\right|z\right>,
	\nonumber
\end{eqnarray}
where we have used the formula
\[
	\left[\hat{a},e^{z\hat{a}^+}\right]_\mp
	=ze^{z\hat{a}^+}.
	\]
Now
\begin{eqnarray}
	\left<1_k\left|\left(1+i\epsilon\hat{A}
	\right)\right|1_l\right>
	&=&
	\delta_{kl}+i\epsilon
	\left<0\left|\hat{a}_k\sum_a A^a \sum_{i,j} T_{ij}^a 
	\hat{a}_i^+ \hat{a}_j \hat{a}_l^+\right|0\right>   \nonumber \\
	&=&
	\delta_{kl}+i\epsilon\sum_a A^a T_{kl}^a
	=(1+i\epsilon A)_{kl}.
	\nonumber
\end{eqnarray}
Also,
\[
	\left<1_k\right|z\left.\right>=z_k.
	\]

\subsubsection{Measure}
The described theory possesses the following
``topological'' gauge symmetry:
\begin{equation}
	\delta  z(x)=\theta(x), \qquad
	\delta\bar{z} (x)=\bar{\theta}(x),
	\label{top.sym.}
\end{equation}
where $\theta (x)$ and $\bar{\theta}(x)$ are
arbitrary except at the boundary $\partial S$
where they vanish. The origin of the symmetry \ref{top.sym.}
will become clear when we convert the action \ref{top.act.}
into a line integral. Integrating by parts in Eq.\ref{top.act.}
and using the Abelian Stokes theorem we obtain
$$
S_{\rm{cl}}=i\oint_{\partial S}
\bar{z}D_iz\,dx^i,
$$
or in a parametrized form
$$
S_{\rm{cl}}=i\oint_{\partial S} \bar{z}
D_t z\,dt  .
$$
To covariantly quantize the theory we shall introduce
the BRS operator $s$. According to the form of the topological gauge
symmetry 
\ref{top.sym.}, the operator $s$ is easily defined by
$$
s\,z =\phi,\qquad s\,\bar{ z}=\bar{\chi}, \qquad
s\phi =0, \qquad s\bar{\chi}=0,
$$
$$
s\bar{\phi}=\bar{\beta}, \qquad s\chi =\beta , \qquad
s\bar{\beta}=0,\qquad s\beta =0,
$$
where $\phi$ and $\bar{\chi}$ are ghost fields in the 
representation $R$, associated to $\theta$ and $\bar{\theta
}$, 
respectively, $\bar{\phi}$ and $\chi$ are the corresponding
antighosts, and $\bar{\beta}$, $\beta$ are Lagrange
multipliers. 
All the fields possess a suitable Grassmann parity
correlated with the parity of $\bar{z}$ and $z$.
Obviously $s^2=0$, and we can gauge fix the action in
Eq.\ref{top.act.} 
in a BRS-invariant manner by simply adding the following
$s$-exact term:
\begin{eqnarray}
	S'&=& s\left( \int_{\cal S} \left( \bar{\phi}\triangle  z
	\pm \bar{ z}\triangle \chi\right) d^2 \sigma \right)
	\nonumber\\	
	&=&
	\int_S \left( \bar{\beta} \triangle z
	\pm \bar{\phi}\triangle \phi \pm \bar{\chi}\triangle\chi
	+\bar{z} \triangle \beta \right) d^2 \sigma.
	\nonumber 
\end{eqnarray}
The upper (lower) sign stands for the fields $\bar{z}$,
$z$ of bosonic (fermionic) statistics. Integration after
the ghost fields yields some numerical factor and
the quantum action
\begin{equation}
	S=S_{\rm{cl}}+\int_S \left(\bar{\beta}
	\triangle z +\bar{z}\triangle\beta\right)
	d^2\sigma.
	\label{Sq}
\end{equation}
If necessary, one can insert $\sqrt{g}$ into the second
term, which is equivalent to  change of variables.
Thus the partition function is given by
$$
Z=\int e^{iS}\,D\bar{z}\,Dz\,D\bar{\beta}\,D\beta ,
$$
with the boundary conditions: $\bar{\beta}|_{\partial S}=
\beta |_{\partial S}=0$. 

One can observe that the job
the fields $\bar{\beta}$ and $\beta$ are supposed to do
consists in eliminating a redundant integration inside
$S$.
The gauge-fixing condition following from Eq.\ref{Sq} imposes
the following constraints 
$$
\triangle  z =0, \qquad \triangle \bar{ z}=0. 
$$
Since  values of the fields $z$ and $\bar{z}$ are
fixed on the boundary $\partial S=C$, we deal
with  the two
well-defined $2$-dimensional Dirichlet problems. 
The solutions of the Dirichlet problems fix values of
$z$ and $\bar z$ inside $S$.
Another, more singular gauge-fixing, is proposed in [Bro92].

The issue of the measure has been also discussed in [HU00].

\subsection{Generalizations}
%
\subsubsection{Topology}
Up to now we have investigated the  non-Abelian Stokes theorem for topologically trivial situation. The term {\it topologically trivial situation} means, in this context, that the loop we are integrating along in the  non-Abelian Stokes theorem is unknotted in the sense of theory of knots [Rol76]. It appears that in contradistinction to the Abelian case, the non-Abelian one is qualitatively different. If the loop $C$ is topologically non-trivial and the bounded surface $S$ ($\partial S=C$) is not simply connected, the parameter space given in the form of a unit square (as in the proof of the  non-Abelian Stokes theorem) is not appropriate. The  non-Abelian Stokes theorem presented in the original form applies only to a surface $S$ homeomorphic to a disk (square). But still, of course, the standard (topologically trivial) version of the  non-Abelian Stokes theorem makes sense locally. What means {\it locally} will appear clear in the due course. The  non-Abelian Stokes theorem for knots (and also for links---multicomponent loops) has been formulated by Hirayama, Kanno, Ueno and Yamokoshi in 1998 [HKUY98]. Interestingly, it follows from this new version of the  non-Abelian Stokes theorem that the value of the line integral along $C$ can be non-trivial (different from 1 even for the field strenght $F_{\mu\nu}(x)$ vanishing everywhere on the surface $S$. It is an interesting result which could have some applications in physics. One can speculate that it could give rise to a new version of the Aharonov-Bohm effect.

To approach the  non-Abelian Stokes theorem for knots we should recall a necessary portion of the standard lore of theory of knots. Since the first task is to find an oriented surface $S$ whose boundary is $C$, we should construct the, so-called, Seifert surface, satisfying the above-mentioned condition by definition. It appears that the Seifert surface for any knot assumes a standard form homeomorphic to a (flat) disk with $2g$ (``thin'') strips attached. The number $g$ is called the genus. The strips may, of course, be horribly twisted and intertwined [Rol76].

\begin{figure}
	\begin{center}	
	\epsfbox{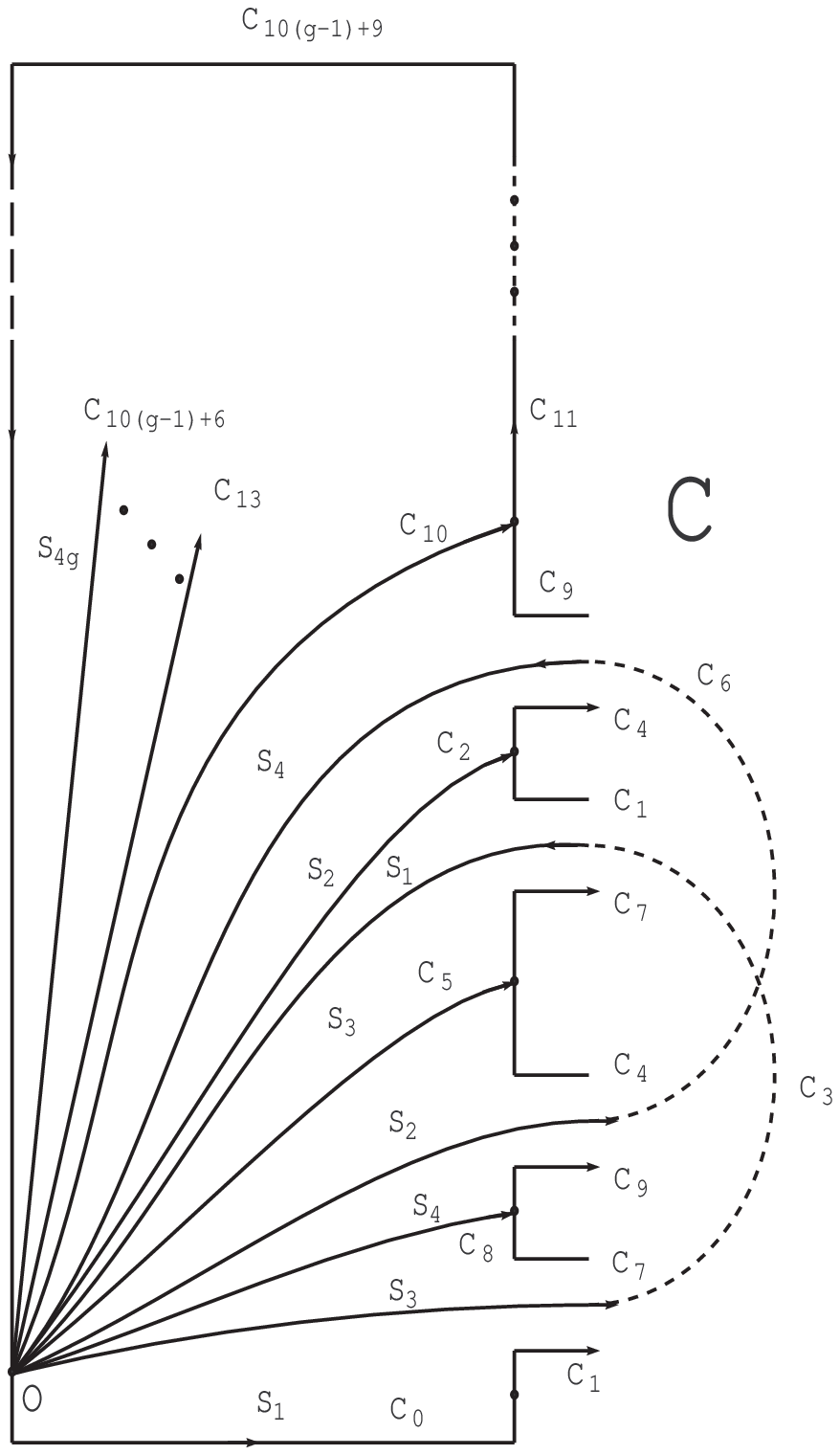}	
	\end{center}	
	\caption{Decomposition of the knotted loop $C$.}
\label{fig:seifert}
\end{figure}

Now we should decompose $C$ and next $S$ into pieces so that one can put from the pieces the slices of that are topologically trivial, and this way they are subject to the standard  non-Abelian Stokes theorem. Such decomposition is shown in Fig.\ref{fig:seifert}.	
Explicitly, it reads
	\begin{eqnarray}
	C&=&\left(C_{10(g-1)+9}\ldots C_{11}\right)
	\left(C_9C_7C_4C_1\right)C_0 	   \nonumber \\
	&=&\left(\prod_{k=0}^{g-1} C_{10k+9} C_{10k+7}
	C_{10k+4}  C_{10k+1} \right) C_0, 
	\qquad {\rm for}\quad g\geq1,
\end{eqnarray}
and
	\[
	C=C_0, \qquad {\rm for}\quad g=0.
	\]
Next
\begin{eqnarray}
	C&=&\Big[C_{10(g-1)+6}
	\underbrace{\big(C_{10(g-1)+6}^{-1}
	\hskip-0.7cm\stackrel{\underbrace{C_{10(g-1)+10}^{-1}}}{}
	C_{10(g-1)+9} C_{10(g-1)+8}\big)}_{S_{4g}}
	C_{10(g-1)+3}\nonumber\\
	 &&\quad\ldots C_{13} \underbrace{\big(C_{13}^{-1}
	C_{12}^{-1}C_{11}C_{10}\big)}_{S_5}\Big]\cdot
	\Big[
	C_6\underbrace{\big(C_6^{-1} C_{10}^{-1} C_9 C_8 
	\big)}_{S_4}   \nonumber \\
	&&\quad
	C_3^{-1}\underbrace{\big(C_3 C_8^{-1} C_7 C_5 
	\big)}_{S_3}
	C_6^{-1}\underbrace{\big(C_6 C_5^{-1} C_4 C_2 
	\big)}_{S_2}C_3\underbrace{\big(C_3^{-1} C_2^{-1}
	C_1 C_0 \big)}_{S_1}
	\Big]	 \nonumber \\
	&=&
	\prod_{k=0}^{g-1} C_{10k+6}\underbrace{%
	\big(C_{10k+6}^{-1}\underline{C_{10k+10}^{-1}}	
	C_{10k+9}C_{10k+8}\big)}_{S_{4k+4}}\nonumber\\
	&&\quad
	C_{10k+3}^{-1} \underbrace{\big(C_{10k+3}C_{10k+8}^{-1}	
	C_{10k+7}C_{10k+5}\big)}_{S_{4k+3}}\nonumber\\
	&&\quad
	C_{10k+6}^{-1} \underbrace{\big(C_{10k+6}C_{10k+5}^{-1}	
	C_{10k+4}C_{10k+2}\big)}_{S_{4k+2}}\nonumber\\
	&&\quad
	C_{10k+3} \underbrace{\big(C_{10k+3}^{-1}C_{10k+2}^{-1}	
	C_{10k+1}C_{10k}\big)}_{S_{4k+1}}\nonumber\\
	&=&
	\prod_{k=0}^{g-1}
	C_{10k+6} S_{4k+4}C_{10k+3}^{-1} S_{4k+3}
	C_{10k+6}^{-1} S_{4k+2}C_{10k+3} S_{4k+1},
	\nonumber
		\end{eqnarray}
	where $C_{10k+10}|_{k=g-1}
	\equiv
	C_{10g}=1$.

Further details are described in the original paper [HKUY98], and continued in [BD01].

\subsubsection{Higher-dimensional forms}
The theorem being considered up to now is a very particular, though seemingly
the most important, non-Abelian version of the Stokes
theorem. It connects a non-Abelian differential one-form in dimension
$1$ and a two-form in dimension $2$.  The forms are of a very
particular shape, namely, the connection $1$-form and the
curvature $2$-form. Now, we would like to discuss possible generalizations to arbitrary, higher-dimensional differential forms in arbitrary
dimensions. Since there may be a lot of variants of such
generalizations depending on a particular mathematical and/or
physical context, we will start from giving a general recipe.

Our idea is very simple. First of all, working in the framework of the path-integral formalism, we should
construct a topological field theory of auxiliary
topological fields on $\partial N$, the boundary of
the $d$-dimensional submanifold $N$, in (an) external (gauge) field(s) we are
interested in. Next, we should quantize the theory, i.e.\
build the partition function in the form of a
path-integral, where auxiliary topological fields are
properly integrated out. Thus, the LHS of the non-Abelian Stokes theorem has been constructed. Applying the Abelian Stokes
theorem to the (effective) action (in the exponent of the
path-integral integrand) we obtain the ``RHS'' of the non-Abelian Stokes
theorem. If we also wish to extend the functional measure
to the whole $N$ we should additionaly quantize the
theory on the RHS to eliminate the redundant functional integration inside $N$.

The example candidate for the topological field theory defining the LHS of the non-Abelian Stokes theorem could be given by the (classical) action
\begin{equation}
	S_{\rm Top}  = \frac{1}{2}\oint_{\partial N} {\left( {\bar zd_A \zeta  + d_A \bar \zeta z + \bar zBz} \right)}, 
	\label{STop}
\end{equation}
where $z$ and $\bar z$ are zero-forms, $\zeta$ and $\bar\zeta$ are ($d-2$)-forms (all the forms are in an irreducible reprezentation $R(G)$), $d_A$ is the exterior covariant derivative
\[
d_A\zeta\equiv d\zeta+A\zeta,
\qquad
d_A\bar\zeta\equiv d\bar\zeta-A^T\bar\zeta.
\]
The non-Abelian $B$-field naturally appears in the context of (topological) gauge theory (see, Eq.\ref{SBF}).
Now, the Abelian Stokes theorem should do.

Generalization of the non-Abelian Stokes theorem to higher-degree forms in the operator language seems more difficult and practically has not been attempted (see, however [Men83] for an introductory discussion of this issue).

\section{Applications}
The number of applications of the  non-Abelian Stokes theorem is not so large as in the case of the Abelian Stokes theorem but neverthelles it is the main motivation to formulate the  non-Abelian Stokes theorem at all. It is interesting to note that in contradistinction to the Abelian Stokes theorem, which formulation is ``homogenous'' (unique), different formulations of the  non-Abelian Stokes theorem are usefull for particular purposes/applications. From a purely techincal point of view, one can classify applications of the  non-Abelian Stokes theorem as exact and approximate. The term {\it exact applications} means that one can perform successfully an ``exact calculus'' to obtain an interesting result, whereas the term {\it approximate application} means that a more or less controllable approximation (typically, perturbative) is involved in the calculus. Since exact applications seem to be more convincing and more illustrative for the subject we will basically confine ourselves only to presentation few of them.

Since the  non-Abelian Stokes theorem applies to non-Abelian gauge theories, and non-Abelian gauge theories are non-linear, it is not so strange that exact applications are scarce. In fact they are limited to low-dimensional cases and/or topological models, which are usually exactly solvable. Our first case considered is pure, two-dimensional ordinary (almost topological)  Yang-Mills gauge theory.
But a rich source of applications of the non-Abelian Stokes
theorem is coming from topological field theory of
Chern-Simons type.  Path-integral procedure gives the
possibility of obtaining skein relations for knot and link
polynomial invariants. In particular, it appears that only the
path-integral version of the non-Abelian Stokes theorem
permits us to nonperturbatively and covariantly generalize
the method of obtaining topological invariants [B90].

As a by-product of our approach we have computed the
parallel-transport operator $U$ in the holomorphic
path-integral representation. In this way,
we have solved the problem of saturation of Lie-algebra
indices in the generators $T^a$. This issue appears, for
example, in the context of equation of motion for
Chern-Simons theory in the presence of Wilson lines (an interesting connection with the
Borel-Weil-Bott theorem and quantum groups has  been also
suggested). Our approach enables us to write those
equations in terms of $\bar{z}$ and $z$ purely classically.
Incidentally, in the presence of Chern-Simons interactions
the auxiliary fields $\bar{z}$ and $z$ acquire fractional
statistics, which could be detected by braiding.  To
determine the braiding matrix one should, in turn, find the
so-called monodromy matrix, e.g. making use of non-Abelian
Stokes theorem.

\subsection{Two-dimensional Yang-Mills theory}
There is a huge literature on the subject of the two-dimensional Yang-Mills theory approaching it from different points of view. One of the latest papers is [AK00], where a list of references to earlier papers is given. Two-dimensional Yang-Mills theory is a specific theory. From the dynamical point of view it is almost trivial---there are no local degrees of freedom, as a standard canonical analysis indicates. In fact it is ``semi-topological'' field theory, i.e.\ roughly it only describes combinatorial/topological phenomena and surface areas.

There are lot of important and interesting aspects in two-dimensional Yang-Mills theory. One of them is the issue of determination of ``physical'' observables---Wilson loops (\ref{TrPexp}). Calculation of the Wilson loops $W_R(C)$ in two-dimensional Yang-Mills theory can be facilitated by the use of the  non-Abelian Stokes theorem.

A nice feature of (Euclidean) two-dimensional Yang-Mills gauge theory defined by the action
\begin{equation}
	S_{\rm 2dYM}(A)
	=\frac{1}{4}\int_M F_{AB}^a(A) F^{a\,AB}(A) \sqrt{g}\,d^2x,
	\qquad
	A,B=1,2,
	\label{YMS}
\end{equation}
is the possibility to recast the action, and next, and more importantly, the whole partition function to the form
\begin{equation}
	Z=\int DF\,e^{-S_{\rm 2dYM}(F)},
	\label{YMZ}
\end{equation}
where now $F$ is an independent field, and the action $S_{\rm 2dYM}$ is of the same form as the original Eq.\ref{YMS} but this time without $A$ dependence.

Let us now consider  ``physical'' observables, i.e.\ Wilson loops. Confronting the partition function \ref{YMZ} with the form of the Wilson loop transformed by the  non-Abelian Stokes theorem to a surface expression \ref{op.NAST} we can see that a kind of a Gaussian functional integral emerges. For an Abelian theory, we would exactly obtain  an easy Gaussian functional integral, but in a non-Abelian case we should be more careful because $\cal F$ is a path-dependent object.
The fact that $\cal F$ is path-dependent can be ignored in the case of a single loop due to the commutativity of the infinitesimal surface integrals (see below). Since, accordingly to the  non-Abelian Stokes theorem
\[
W_R(C)={\rm Tr}_R{\cal P}\exp{\left(i\int_S
{\cal F}\sqrt{g}\,d^2x\right)},
\]
where ${\cal F=F}_{12}$. For the expectation value
\[
\left<W_R(C)\right>
=Z^{-1}\int DF\exp{\left(-S_{\rm 2dYM}\right)}W_R(C),
\]
we obtain ([Bro90])
\begin{eqnarray}
\left<W_R(C)\right>
&\propto&
\int DF\exp{\left(-\frac{1}{2}\int_M F^aF^a
\sqrt{g}\,d^2x\right)}
{\rm Tr}_R{\cal P}\exp{\left(i\int_S
F\sqrt{g}\,d^2x\right)}
\nonumber
\\
&=&
\int \prod_{x\in M\backslash S}
dF(x)\exp{\left(-\frac{1}{2}\int_{M\backslash S}
F^aF^a\sqrt{g}\,d^2x\right)}
\nonumber
\\
&&
\cdot{\rm Tr}_R{\cal P}\int \prod_{x\in S}dF(x)
\exp{\left[i\int_S \left(-\frac{1}{2}F^aF^a
+iF^aT^a\right)\sqrt{g}\,d^2x\right]}
\nonumber
\\
&\propto&
{\rm Tr}_R
\exp{\left(-\frac{1}{2}T^aT^a\int_S 
\sqrt{g}\,d^2x\right)}.
\nonumber
\end{eqnarray}
Thus, finally
\[
\left<W_R(C)\right>
={\rm Tr}_R\exp{\left[-\frac{1}{2}C_2(R)S\right]}
=\dim{R}\exp{\left[-\frac{1}{2}C_2(R)S\right]},
\]
where
\[
S=\int_S\sqrt{g}\,d^2x,
\]
and
\[
T^aT^a=C_2(R),\qquad
{\rm Tr}_RI=\dim{R}.
\]
In the case of $n$ non-overlapping regions $\{S_i\}$, $i=1,\ldots,n$, $C_i=\partial S_i$,
$S_i\cap S_j=\emptyset$ for $i\neq j$, and $n$ irreducible representations $R_i$ of the group $G$ with the generators $T_i$, we immediately obtain, literally repeating the last derivation, the formula for the expectation value of the product of the $n$ Wilson loops
\[
\left<\prod_{i=1}^n W_{R_i}(C_i)\right>
=\prod_{i=1}^n\dim{R_i}
\exp{\left[-\frac{1}{2}C_2(R_i)S_i\right]}.
\]
The case of the overlapping regions $\{S_i\}$ is a bit more complicated [Bro90]. First of all, one has to decompose the union of all regions $\{S_i\}$, $\partial S_i=C_i$ into a disjoint union of connected, i.e.\ not intersected by the loops, regions $\{S_\alpha\}$. Each loop $C_i$ is next deformed into an equivalent loop $C_i'$, which is a product of ``big'' (not infinitesimal) lassos independently (a lasso per a region) covering each connected region $S_{\alpha_i}$, $S_{\alpha_i}=S_\alpha\cap S_i$ ($S_{\alpha_i}\in \{S_\alpha\}$). The lassos coming from the different loops $C_i'$ but covering the same connected region $S_\alpha$ should necessarily be arranged in such a way to enter the region $S_\alpha$ at the same base point $O_\alpha$. Consequently, the connected region $S_\alpha$, $S_\alpha\subset S_{i_1}\cap\ldots\cap S_{i_k}$, $i\leq k\leq n$, can be covered with the $k$ identical copies of the net of ``small'' (infinitesimal) lassos. Every Gaussian functional integration with respect to the infinitesimal area $\delta S$ enclosed by an infinitesimal lasso can be easily performed yielding
\begin{equation}
	\exp{\left(-\frac{1}{2}\delta S\,T_\alpha^2\right)},
	\label{YMexp}
\end{equation}
where
\[
T_\alpha=\sum_{i\in\alpha} T_i,
\qquad
T_i=I\otimes\dots\otimes T_i\otimes\ldots\otimes I.
\]
Integration with respect to the consecutive infinitesimal areas gives the terms of the form \ref{YMexp}. Since $T_\alpha$ is a generator of $\bf g$ in a product representation $R_\alpha$, $R_\alpha=R_{i_1}\otimes\ldots\otimes R_{i_k}$, $T_\alpha^2$ is a Casimir operator. Accordingly, \ref{YMexp} commutes with the product of the parallel-transport operators acting in the product representation $R_\alpha$. Since the products in the pairs connect every infinitesimal area $\delta S$ with the base point $O_\alpha$, they cancel each other. This fact means that the integral with respect to the whole region $S_\alpha$ is given only by the infinite product of the terms \ref{YMexp} and reads
\begin{equation}
	M_\alpha=\exp{\left(
	-\frac{1}{2}S_\alpha T_\alpha^2\right)}.
	\label{YMM}
\end{equation}
The full expectation value of the $n$ loops $\{C_i\}$ consists of the trace of a product of $M_\alpha$-blocks \ref{YMM} joined with the parallel-transport operators, which are remnants of the primary decomposition of the loops. These joining curves enclose zero areas, and can be deformed into points (without destroying $M_\alpha$-blocks) giving some ``linking'' operators $L_\alpha$. An operator $L_\alpha$ is of a very simple form, namely it is a product of the Kronecker deltas, which contract indices belonging to the same representation but to different $M$'s. Thus, $L$ causes that the matrix multiplication of $M$'s is performed in a prescribed order in each representation sector independently. In other words, $M$ mixes, with some weights, indices of different representations (braiding), whereas $L$ sets the order of the matrix multiplications in a representation sector. $M$ depends on the metric (area of $S$) and group-theoretic quantities, whilts the concrete form of $L$ depends on the topology of the overlaps. Thus, the expectation value of the product of the $n$ Wilson loops is finally given by
\[
\left<\prod_{i=1}^n W_{R_i}(C_i)\right>
=\prod_\alpha L_\alpha M_\alpha.
\]
This analysis is a bit simplified and shortened but gives the flavor of the power of the  non-Abelian Stokes theorem in practical instances.

\subsection{Three-dimensional topological quantum field theory}
Topological quantum field theory has recently
become a fascinating and fashionable subject in
mathematical physics. At present, main 
applications of topological field theory are in mathematics (topology
of low-dimensional manifolds) rather than in physics.
Its application to the issue of classification of knots and links is
one of the most interesting.
To approach this problem one usually tries to somehow
encode the topology of a knot/link . As was first noted by Witten [Wit89], the
problem can be attacked by means of standard
theoretical physics techniques of quantum field theory. In
particular, using three-dimensional Chern-Simons gauge theory one can derive not only all the well-known
polynomial invariants of knots and links but many
of their generalizations as well.
Most authors working in the topological field theory description of polynomial
invariants follow Witten's original approach, which relies
heavily on the underlying conformal-field-theory structure.
There is also a genuinely three-dimensional covariant
approach advocated in its perturbative version in [Smo89] and [CGM90] using the  non-Abelian Stokes theorem in its operator formulation.
We shall sketch an application of the  non-Abelian Stokes theorem to a genuinely three-dimensional, non-perturbative, covariant
path-integral approach to {\it polynomial invariants} of
knots and links in the framework of (topological) quantum
Chern-Simons gauge field theory.

To begin with, we introduce the classical topological Chern-Simons action
on the three-dimensional sphere $S^3$
\begin{eqnarray}
	S_{\rm CS}
	&=&
	{k\over4\pi}\int_{S^3}{\rm Tr}\left(A\wedge
	dA+{2\over3}A\wedge A\wedge
	A\right)
	\nonumber\\
	&=&
	{k\over4\pi}\int_{S^3}d^3x\,
	\varepsilon^{ijk}{\rm Tr}\left(A_i\partial_j
	A_k+{2\over3}A_i A_j A_k\right),
	\label{CSS}
\end{eqnarray}
where $k\in{\bf Z}^\pm$.
The use of Eq.\ref{CSS} is not obligatory. 
One could as well
choose the action of the so-called $BF$-theory
\begin{equation}
	S_{\rm BF}={k\over4\pi}\int_{{S}^3}d^3x\,
	\varepsilon^{ijk}{\rm Tr}(B_i F_{jk}),
	\label{SBF}
\end{equation}
where $B_i=B_i^a(x)T^a$ is an auxiliary gauge
field, and now $k\in{\bf R}^\pm$.

To encode the topology of a link ${\cal L}=\{C_i\}$ into
a path integral we introduce an
auxiliary one-dimensional topological
field theory (topological quantum mechanics) in an external
gauge field $A$, living on the corresponding loop
$C_i$.  The classical action of this theory
is chosen in the form (see, Eq.\ref{clas.L})
\begin{equation}
	S_{C_i}(A)
	=i\oint_{C_i}dt\,\bar{z}_iD_t z_i,
	\label{CSo}
\end{equation}
where the multiplet of scalar fields
$\bar{z}_i$, $z_i$ transforms in a irreducible
representation $R_i$. The partition function corresponding
to \ref{CSo} has the following standard form
$$
Z_i(A)=\int D^2z_i\exp\left[iS_{C_i}(A)\right].
$$
It is obvious that the observables $S_{C_i}(A)$
are akin to the Wilson loops.

We define the topological invariant of the link ${\cal
L}$ as the (normalized) expectation value
\begin{equation}
	\left<\prod_iZ_i(A)\right>
	\equiv\left[\int d\mu\exp(iS)\right]^{-1}
	\int d\mu\exp(iS)\prod_iZ_i(A),
	\label{CSvac}
\end{equation}
where $S$ consists of $S_{\rm CS}$ plus quantum terms, and the measure should also contain the auxiliary fields.
We can calculate \ref{CSvac} recursively, using the so-called
{\it skein relations}. Thus, our present task reduces to the
derivation of the corresponding skein relation. To this
end, we  consider a pair of loops, say $C_1$ and $C_2$, where a part of $C_1$, forming a small loop
$\ell$ ($\ell=\partial N$), is wrapped round $C_2$. In other words, $C_2$ pierces $N$ at a point $P$.
\placedrawing{skein.fig}{$C_1$ and $C_2$, where a part of $C_1$, forming a small loop
$\ell$ ($\ell=\partial N$), is wrapped round $C_2$.}{fig:skein}
Such an
arrangement can be interpreted as a preliminary step
towards finding the corresponding monodromy matrix {\bf M}.
Having given the loop $\ell$ we can utilize the  non-Abelian Stokes theorem (actually the Abelian Stokes theorem for \ref{CSo}) in its holomorphic version, obtaining Eq.\ref{top.act.}.
In general position, $N$ and $C_2$ can
intersect in a finite number of points, and the
contribution to the path-integral coming from these points
can be explicitly calculated.  We 
replace the curvature in \ref{top.act.} by the functional
derivative operator
$$
F_{ij}^a(x)\longrightarrow{4\pi\over
ik}\varepsilon_{ijk}{\delta\over\delta
A_k^a(x)}.
$$
This substitution yields an equivalent expression provided the order of terms in \ref{CSvac} is such
that the functional derivative can act on $S_{\rm CS}$
producing $F$. Using formal
translational invariance of the product measure $DA$, and
functionally integrating by parts in \ref{CSvac} with respect
to $A$, we obtain, for each intersection point $P$, the {\it monodromy operator}
\begin{equation}
	M=\exp\left[{4\pi\over ik}(\bar z_1T_1^a z_1)
	(\bar z_2 T_2^a z_2)(P)\right].
	\label{CSM}
\end{equation}
To calculate the {\it matrix elements} of \ref{CSM} one
utilizes the following scalar product
$$
(f,g)={1\over2\pi i}\int
fg\exp(\bar z z)d\bar z dz.
$$
This kind of the scalar product is implicit in our derivations of the path integral.
Expanding \ref{CSM} in a power series, multiplying with respect to
this scalar product, and re-summing, we get the {\it
monodromy matrix}
$$
{\bf M}=(\bar z_1\bar z_2, M\,z_2z_1)
=\exp\left({4\pi\over ik}T_1^a\otimes T_2^a\right).
$$
The square root of the monodromy matrix gives rise to the, so-called, braiding matrix $\bf B$ resposible for a proper form of skein relations, yielding knot/link invariants.

\subsection{Other applications}
We could continue the idea of the previous section and try to generalize it to higher-dimensional (topological) theories. To this end we should use a generalization of the non-Abelian Stokes theorem to non-Abelian forms of higher degree, e.g.\ following the approach proposed in Eq.\ref{STop}, and yielding the resluts obtained in [Bro94].

Quite a different story is the possibility to apply the non-Abelian Stokes theorem (in the coherent-state version) to computations in QCD (QCD string, area-low, etc.) [AE97], [DP95], [Sim92] and gravity [DP01]. Such calculations are usually posssible only perturbatively, since their results as not rigorously controlable are uncertain. Concerning coordinate-gauge approach and field strength and dual formalism, extended lattice formulation for spin systems has been proposed in [BH84-1], and for gauge theories in [BH84-2].

%
\section{Summary}
In this short review we have addressed main issues related to the non-Abelian Stokes theorem. The two principal approaches (operator and path-integral) to the non-Abelian Stokes theorem have been formulated in their simplest possible forms. A generalization for a knotted loop as well as a suggestion concerning higher-degree forms have been also presented. Only non-perturbative applications of the non-Abelian Stokes theorem (to low-dimensional gauge theories) have been described. The review is not comprehensive and is rather directed towards topological aspects reflecting author's interests.

%
\paragraph{Acknowledgments}
The author is grateful to
Professors M. Blau and P. Kosi\'nski for interesting discussions at very early stages of the development of some of these ideas. The author would also like to thank Professors D.~Diakonov, M.~B.~Halpern and F.~Mansouri for their comments.
The work has been supported by the grant of the University of \L\'od\'z.

%

\end{document}